\documentclass[final,5p,times,twocolumn,sort,compress]{elsarticle}

\usepackage{graphicx}

\usepackage{amsmath,amsfonts,amssymb,amsxtra}
\usepackage{units,formula}
\usepackage{sidecap}

\makeatletter
\DeclareRobustCommand{\chemical}[1]{%
  {\(\m@th
   \edef\resetfontdimens{\noexpand\)%
       \fontdimen16\textfont2=\the\fontdimen16\textfont2
       \fontdimen17\textfont2=\the\fontdimen17\textfont2\relax}%
   \fontdimen16\textfont2=2.7pt \fontdimen17\textfont2=2.7pt
   \mathrm{#1}%
   \resetfontdimens}}
\DeclareRobustCommand{\bchemical}[1]{%
  {\(\m@th
   \edef\resetfontdimens{\noexpand\)%
       \fontdimen16\textfont2=\the\fontdimen16\textfont2
       \fontdimen17\textfont2=\the\fontdimen17\textfont2\relax}%
   \fontdimen16\textfont2=2.7pt \fontdimen17\textfont2=2.7pt
   \mathbf{#1}%
   \resetfontdimens}}
\makeatother

\biboptions{sort,compress}

\journal{Physica C}

\newcommand{\nsmovz}{\chemical{Nd_{0.33}Sr_{1.67}MnO_4}}

\newcommand{\pcmovz}{\chemical{Pr_{0.33}Ca_{1.67}MnO_4}}

\newcommand{\lsmov}{\chemical{La_{1-x}Sr_{1+x}MnO_4}}
\newcommand{\lsmohd}{\chemical{La_{0.5}Sr_{1.5}MnO_4}}
\newcommand{\lsmovo}{\chemical{La_{0.42}Sr_{1.58}MnO_4}}
\newcommand{\lsmhd}{\chemical{La_{0.5}Sr_{1.5}MnO_4}}

\newcommand{\lscoz}{\chemical{La_{1.67}Sr_{0.33}CoO_4}}

\newcommand{\lscoo}{\chemical{La_{2-x}Sr_{x}CoO_4}}
\newcommand{\lscov}{\chemical{La_{2-x}Sr_{x}CoO_4}}
\newcommand{\lcoo}{\chemical{La_2CoO_4}}
\newcommand{\lscopd}{\chemical{La_{1.7}Sr_{0.3}CoO_4}}

\newcommand{\lscopf}{\chemical{La_{1.5}Sr_{0.5}CoO_4}}

\newcommand{\lccopf}{\chemical{La_{1.5}Ca_{0.5}CoO_4}}

\newcommand{\lsco}{\chemical{La_{2-x}Sr_{x}CuO_4}}
\newcommand{\lco}{\chemical{La_{2}CuO_4}}

\newcommand{\lsno}{\chemical{La_{2-x}Sr_{x}NiO_4}}
\newcommand{\lsnod}{\chemical{La_{2-x}Sr_{x}NiO_{4+\delta}}}
\newcommand{\lsnodd}{\chemical{La_{1.67}Sr_{0.33}NiO_4}}
\newcommand{\lno}{\chemical{La_{2}NiO_4}}
\newcommand{\lsnopz}{\chemical{La_{1.8}Sr_{0.2}NiO_4}}

\newcommand{\mnd}{Mn$^{3+}$}
\newcommand{\mnv}{Mn$^{4+}$}
\newcommand{\mb}{$\mu B$}

\newcommand{\vech}{${\boldsymbol\eps}_{ch}$}
\newcommand{\ves}{${\boldsymbol\eps}_S$}
\newcommand{\veo}{${\boldsymbol\eps}_{oo}$}
\newcommand{\grad}{$^{\circ}$}

\begin{document}

\begin{frontmatter}



\title{Neutron scattering studies on stripe phases in non-cuprate materials}


\author[a]{Holger Ulbrich}

\author[a]{Markus Braden}
\ead{braden@ph2.uni-koeln.de}
\address[a]{II.~Physik. Institut, Universit\"at zu K\"oln,
Z\"ulpicher Str.~77, D-50937 K\"oln, Germany}

\begin{abstract}

Several non-cuprates layered transition-metal oxides exhibit clear
evidence for stripe ordering of charges and magnetic moments.
Therefore, stripe order should be considered as the typical
consequence of doping a Mott insulator, but only in cuprates
stripe order or fluctuating stripes coexist with metallic
properties. A linear relationship between the charge concentration
and the incommensurate structural and magnetic modulations can be
considered as the finger print of stripe ordering with localized
degrees of freedom. In nickelates and in cobaltates with
K$_2$NiF$_4$ structure, doping suppresses the nearest-neighbor
antiferromagnetism and induces stripe order. The higher amount of
doping needed to induce stripe phases in these non-cuprates series
can be attributed to reduced charge mobility. Also manganites
exhibit clear evidence for stripe phases with further enhanced
complexity, because orbital degrees of freedom are involved.
Orbital ordering is the key element of stripe order in manganites
since it is associated with the strongest structural distortion
and with the perfectly fulfilled relation between doping and
incommensurability. Magnetic excitations in insulating stripe
phases exhibit strong similarity with those in the cuprates, but
only for sufficiently short magnetic correlation lengths
reflecting well-defined magnetic stripes that are only loosely
coupled.

\end{abstract}

\begin{keyword}


\end{keyword}

\end{frontmatter}


\section{Introduction}
\label{intro}

The discovery of stripe-like magnetic and charge order in
Nd-codoped \lsco \ \cite{1} had a strong impact on the
understanding about doping a Mott insulator. Several theoretical
papers dealt with the introduction of charges into the Hubbard
model and discussed the possibility of holes condensing into
domain walls of the antiferromagnetic ordering \cite{2,3,4,5,5a}.
Charged domain walls were found to be stable solutions of the
still simple models even without taking an electron-phonon
coupling into account \cite{2,3,4,5,5a}. The experimental
confirmation of stripes in some cuprates materials inspired
speculations about the role fluctuating charge inhomogeneity - or
more specifically fluctuating stripes - could play in the
superconducting pairing mechanism \cite{6,7} but this question
still remains matter of controversy. In the cuprates, static
stripe order competes with superconductivity, since the
superconducting transition temperatures and the condensation
energy are strongly suppressed in materials exhibiting static
stripe order \cite{8,9,10,11}. However, fluctuating stripe order
can coexist with stable high-temperature superconductivity and be
involved in the pairing \cite{7}. In several cuprates, the
dispersion of magnetic excitations shows a very characteristic
hourglass shape \cite{12,13,14,15,16,17} which can easily be
explained by magnetic stripe ordering or by slowly fluctuating
stripes. However, a successful description of the magnetic
fluctuations is also obtained by the opposite scenario with
itinerant charge carriers \cite{18}.

\begin{figure*} \centering
{\includegraphics[width=0.999\textwidth]{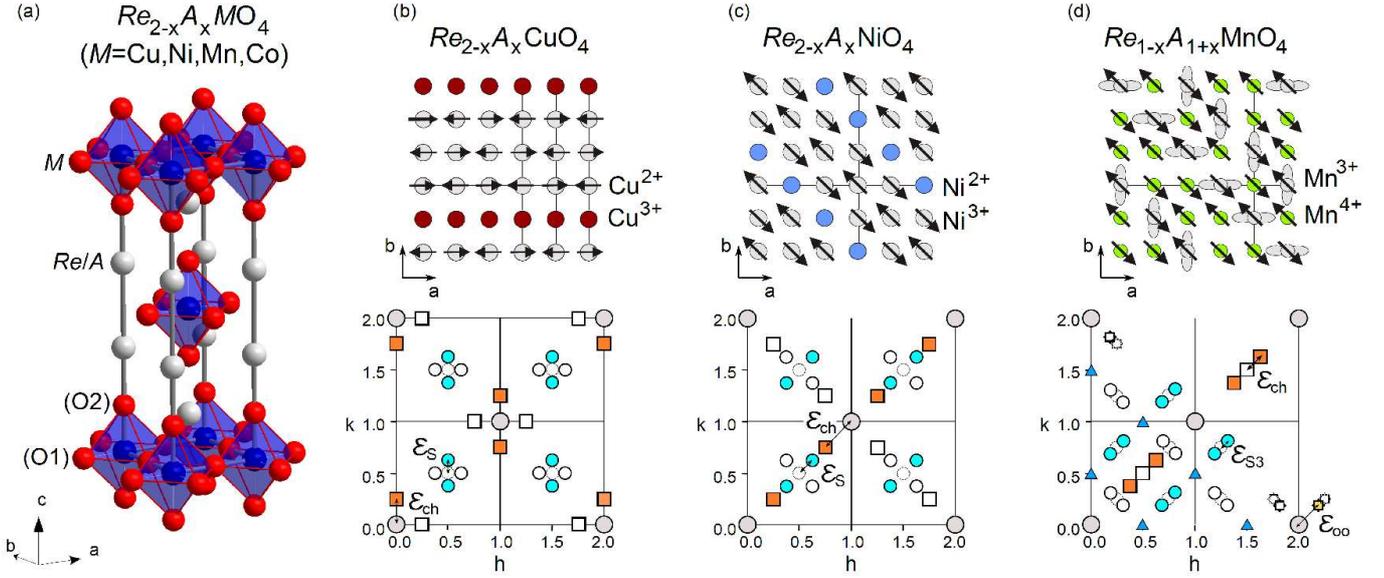} \caption{
Crystal structure of {\it Re}$_{2-x}${\it A}$_x$$M$O$_4$ with {\it
Re}= La or a rare earth, {\it A} = Sr or Ca and {\it M} = Cu, Ni,
Co, Mn. (Upper panels) Real-space drawings (ab-plane) of the
schematic ordering of charges, magnetic moments and orbitals
proposed for cuprates (b), nickelates (c) and manganites(d). In
cuprates and nickelates regions of nn antiferromagnetism are
separated by the charged stripes which act as domain walls. In
layered manganites it is important to take the orbital degree of
freedom at the nominal \mnd \ site into account, because this
mediates the strongest ferromagnetic interaction. (Lower panels)
The reciprocal space pictures for the purely two-dimensional
ordering depicted above. For cuprates and nickelates charge and
magnetic modulation yield corresponding scattering at satellites
displaced from the original Bragg positions by \vech \ and \ves ,
respectively. In the stripe ordering in layered manganites (c)
charges, orbital and magnetic moments give rise to four different
types of super-structure scattering associated with charge,
orbital, \mnd - and \mnv -moment ordering.}} \label{fig1}
\end{figure*}

In order to better understand the role of the stripe instability
in the physics of high-temperature superconductivity it seems most
interesting to investigate stripes phases in related compounds. By
now, stripe-like charge and magnetic order were reported in
several other transition-metal oxides with layered \cite{19,20,21}
or pseudo-cubic structure \cite{22,23,25}. Fig. \ref{fig1}
illustrates the real-space ordering of the different stripe phases
discussed for these materials together with the indication of the
positions of superstructure reflections arising from magnetic and
electronic ordering. These schemes which are frequently discussed
in the literature, however, largely oversimplify the true
ordering. Charge ordering like in most transition-metal oxides
does not correspond to an integer modulation of the electron count
at the metal site, and the charge modulation is not limited to the
metal site \cite{coey} but most of the varying hole density sits
on the bonds and ligands surrounding a metal site. Nevertheless
these simple pictures give the clearest insight to the sometimes
complex ordering schemes.

In the cuprates with large doping, charges are found to segregate
in stripes running along the Cu-O bonds and are therefore called
horizontal (or vertical) stripes \cite{tranquada-review}. In the
nickelates stripes are observed to run along the diagonals
(diagonal stripes) as it is also reported for low-doped \lsco
 \ \cite{diag-lsco,diag-lsco1,diag-lsco2,diag-lsco3,diag-lsco4,diag-lsco5}. The regular stacking of the stripes gives rise to
superstructure-reflection intensities displaced by a propagation
vector \vech \ from the Bragg peaks. Throughout the paper we give
reciprocal space vectors in reduced lattice units referring to the
small perovskite mesh with a lattice constant of $\sim$ 3.8 \AA .
The commensurate arrangements in Fig. \ref{fig1} explains that the
in-plane length of \vech \ must be proportional to 1/n with $n$
the stripe distance given in d-spacings of the corresponding
planes. If the average stripe distance is incommensurate the
resulting superstructure is incommensurate as well. In real space
this just means that the stripe distance varies on the local scale
and that only the mean stripe distance is defined by the doping.
This arrangement can be considered as a soliton lattice. Since the
stripe distance in this simple model is just given by the amount
of additional charges, it must amount to  $n = 1/n_h$ times the
d-spacing of the lattice with $n_h$ the charge concentration. The
length of \vech \ itself is then proportional to $n_h$. In
stoichiometric \lsno \ we thus expect \vech $ = (x, x, q_l)$. In
the horizontal or vertical stripes in the cuprates, one hole seems
to occupy two sites and one gets \vech $ = (0.5\cdot n_h, 0,
q_l)$. In most cases, the fundamental structure is tetragonal and,
therefore, unable to pin the stripe direction along a certain
direction, or stripes in neighboring planes run in perpendicular
direction. Therefore, the low-temperature scattering pattern
consists of the superposition of at least two domain orientations
with stripes running along perpendicular directions. Neutron
diffraction is not directly sensitive to the electronic charges.
However, a charge modulation on the metal site implies a
modulation of the bond distances via the well-known relations
between bond length and bond strength. This bond-length modulation
can be analyzed through the displacements of the oxygen atoms.
Since neutron scattering is more sensitive on the oxygen than
x-rays, the precise determination of the charge-order related
distortion can be more easily performed with neutron diffraction
techniques \cite{bacon}. Direct insight can, however, be obtained
when using resonant x-ray diffraction in particular in the soft
energy range \cite{abamonte}. Neutron diffraction thus senses the
lattice distortions generated by the charge ordering. Since this
is not a scalar scattering contribution the charge-order related
intensities can only be observed at positions with finite
structure factor. For the charge ordering we essentially expect a
breathing-type distortion which in general can be studied only in
an at least partially longitudinal configuration
\cite{bradenphon1,bradenphon2}.

As the stripe acts as a domain wall for the antiferromagnetic
order between two charge stripes, the wave length of the magnetic
modulation is twice as large as the structural one. Magnetic
super-structure intensities appear thus at positions shifted by
\ves \ from the magnetic Bragg peak of the commensurate
nearest-neighbor antiferromagnetic order at (0.5, 0.5, $q_l$), and
the in-plane length of \ves \ is just half that of \vech: $\eps_S$
=$\frac{1}{2}\cdot\eps_{co}$. This resembles the situation in a
classical spin-density wave (SDW) like that in Cr where a coupling
with a structural modulation with half period arises from the
magnetostriction \cite{cr1,cr2,cr3}. The crystal structure
distorts in the way, that the magnetic interaction is enhanced in
regions of large moments and reduced in regions of low moments.
The doubling of the nuclear modulation vector with respect to the
magnetic one, arises from the fact that the magnetostriction does
not depend on the sign of the moments and that the real-space wave
length is therefore half as long. The coupled charge and magnetic
order in the cuprates and in the nickelates as well as the charge
orbital and magnetic ordering in the manganites can in principle
be interpreted in such SDW and charge-density wave (CDW) model,
but the larger amount of experiments point to the local stripe
model at least for the insulating materials discussed here. In
particular the clear relation between the incommensurability and
the charge concentrations, which is well fulfilled in nickelates,
cobaltates and manganites, is difficult to explain with a SDW-CDW
model basing on a Fermi-surface instability.

The diagonal stripe pattern shown in Fig. \ref{fig1}(c) applies to
the nickelates and most likely also to the layered cobaltates
\cite{20} and low-doped cuprates
\cite{diag-lsco,diag-lsco1,diag-lsco2,diag-lsco3,diag-lsco4,diag-lsco5}.
Stripe ordering in manganites was first reported for the
pseudo-cubic perovskite phases \cite{22,23,25}. These materials
are, however, heavily twinned rendering the interpretation of any
neutron scattering experiments very difficult. Therefore, a fully
consistent picture could only be established for a layered
manganite, \lsmovo , which was slightly overdoped in respect to
the stable ordering at half doping \cite{21}. The orbital degree
of freedom plays an important role in the ordering processes in
the manganites, whereas it is fully suppressed in the cuprates.
This considerably enhances the complexity of the stripe order in
the manganites, see Fig. \ref{fig1}(d). Another difference
concerns the doping range where stripe patterns are observed. In
the cuprates stripes appear at rather low doping and thus close to
the simple nn antiferromagnetic phase. In nickelates and
cobaltates the stripe magnetic ordering is seen already at higher
doping but it is still possible to associate the stripe modulation
with the nn antiferromagnetism in the parent compounds. In the
layered manganites one typically considers LaSrMnO$_4$ with only
\mnd \ (with a $3d^4$ electronic configuration) as the starting
material. Doping leads to the suppression of the nn AFM order in
LaSrMnO$_4$ and to a quite stable charge, orbital, and magnetic
order at half doping, \lsmhd , where orbital order seems to be
essential for the magnetic interaction. In this phase
ferromagnetic zigzag chains are coupled antiferromagnetically;
this magnetic arrangement is called CE-type. Stripe-like ordering
of the various degrees of freedom develops in the manganites only
in relation with the CE-type order at half doping, see Fig.
\ref{fig1}(d).

\begin{figure} \centering {
\includegraphics[width=0.75\columnwidth]{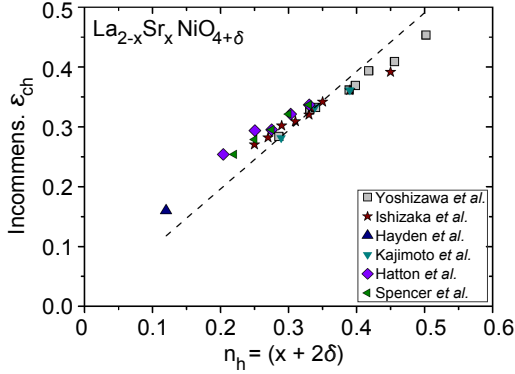}
\caption{Dependence of the incommensurate modulation of spin and
nuclear structure on the hole concentration $n_h = x+2\cdot\delta$
in \lsnod , data are taken from \cite{n12,n2,n15,n15p,nxr2,nxr3}
and correspond to the low-temperature values. Note that the
incommensurability is not exactly following $\eps _{ch}$ = $n_h$
(indicated by the dashed line), but the incommensurability tends
to approach $\eps _{ch}$$\sim$$\frac{1}{3}$. } \label{fig2}}
\end{figure}

In this article we will resume the overwhelming amount of work
performed on the stripe phases in nickelates and compare the
results with the properties of the more recently reported stripe
ordering in layered cobaltates and manganites.

\section{Stripe order of charges and magnetic moments in layered nickelates}
\label{nickelates}

After the discovery of superconductivity in the cuprates, the
electronic properties of layered nickelates were studied as well,
but metallic behavior can be induced only for very high amounts of
doping, $x\sim$1, in \lsno \ \cite{takeda,cava}. The low
conductivity in \lsno \ at moderate doping is attributed to the
formation of small polarons \cite{ng2,ng2p}. Besides the interest
in the the stripe ordering of charges and electrons discussed
here, doped nickelates attract strong attention due to the
observation of an electric-field driven brake down of the
insulating state \cite{yamanouchi}, due to a dielectric anomaly
reflecting the glassy electronic state \cite{park1}, and most
recently due to the
 hole Fermi surface observed for a $d_{x^2-y^2}$ band in highly doped materials
\cite{uchida}.

Pure \lno \ can be doped either by Sr or Ca substitution or by
insertion of excess oxygen leading to \lsnod \
\cite{takeda,cava,ng1}. The precise control of the oxygen
stoichiometry is particularly difficult during the growth of
single crystals strongly affecting the reliability of the absolute
value of the total charge concentration, $n_h = x+2\delta$ in
\lsnod . Synthesizing stoichiometric \lsno \ becomes more
difficult with decreasing Sr content, since these compounds easily
incorporate large amounts of excess oxygen \cite{praba}.

Pure \lno \ exhibits a structural phase transition from the
high-temperature tetragonal (HTT) phase to a low-temperature
orthorhombic (LTO) phase \cite{ng1p} like pure \lco , but due to
higher bond-length mismatch the transition occurs at higher
temperature in \lno , T$_{LTO}\sim$781\ K \cite{friedt}. At 70\ K
a second structural phase transition occurs into a low-temperature
tetragonal (LTT) phase \cite{n1}. The LTO and LTT phases are both
characterized by tilting of the NiO$_6$ octahedra but the tilt
axes differ: In the LTO phase the tilt axis is along the diagonals
of the small perovskite (or parallel to NiO$_6$ octahedron edge),
whereas octahedra tilt around an axis parallel to the bond in the
LTT phase. Monoclinic phases with the tilt axis oriented near but
not exactly along the bonds are labelled low-temperature less
orthorhombic (LTLO).  The structural distortion is suppressed by
the Sr doping \cite{huecker2004,friedt} as it is expected by the
bond-length mismatch model due to the large ionic radius of Sr and
due to the oxidation of Ni similar to the case of \lsco \
\cite{braden1,radaelli}. \lno \ exhibits nn antiferromagnetic
order below T$_N\sim$650\ K with an ordered moment of 1.6 \mb \
\cite{n1} in reasonable agreement with the expectation for S = 1
Ni$^{2+}$ with $3d^8$ configuration.

\begin{figure} \centering {
\includegraphics[width=0.75\columnwidth]{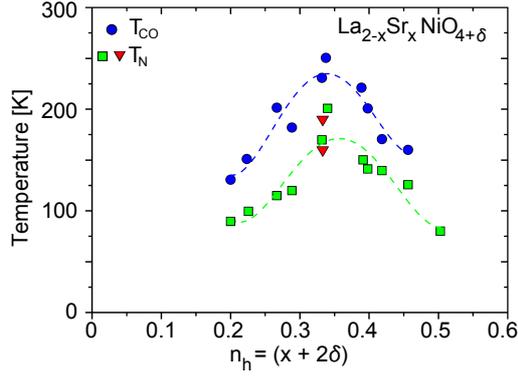}
\caption{Transition temperatures into the charge (T$_{CO}$) and
magnetic (T$_N$) ordered states plotted against the hole
concentration $n_h = x+2\cdot\delta$, data from \cite{n12}. The
red triangles indicate neutron and NMR measurement of $T_N$ of the
same sample indicating that magnetic order is not fully static
below the $T_N$ observed in neutron scattering experiments but
fluctuating with a slow time scale \cite{lee1,nmr}.} \label{fig3}}
\end{figure}

First evidence for an incommensurate magnetic correlation was
obtained in an elastic neutron-scattering study on \lsnopz \ and
antiphase domain walls were discussed as one possible explanation
\cite{n2}. This crystal was annealed at highly reducing conditions
resulting in an oxygen content of 3.96(6). Furthermore, electron
diffraction experiments on several \lsno \ crystals revealed
quasi-2D superlattice peaks whose positions strongly depend on the
Sr concentration \cite{n3}. A study of the x-ray scattering on
comparable crystals revealed strong diffuse contributions at
symmetric satellite positions which, however, cannot be explained
with the charge correlations and may result from thermal diffuse
scattering or some La Sr ordering \cite{n3p}. For an oxygen-doped
crystal of La$_2$NiO$_{4.125}$, Yamada et al. find incommensurate
magnetic scattering to appear at rather high temperature in view
of the large amount of doping \cite{n4p}. For the same material
Tranquada et al. established for the first time the simultaneous
ordering of holes and spins according to the pattern shown in Fig.
\ref{fig1}(c) which precisely corresponds to the hole
concentration of $n_h$ = 0.25 in this material \cite{n4}. The
excess oxygen in La$_2$NiO$_{4+\delta}$ orders upon cooling giving
rise to very complex phase diagrams and to a variety of
superstructure reflections which superpose or overlap with those
associated with the charge and spin stripe ordering
\cite{n5,n6,n6p}. Magnetic moments in La$_2$NiO$_{4.125}$ align
perpendicular to the modulation, i.e. parallel to the stripes, and
are pretty large, about 80\% of the ordered moment in pure \lno \
\cite{n6p}. A more stable charge order was detected for higher
amount of excess oxygen similar to the later observation on \lsno
\cite{n6pp,n10} and allowed for a better characterization of the
underlying structural distortion. The charge incommensurability is
not constant in this material but discontinuously varies upon
cooling passing through several commensurate lock-in values as in
a typical incommensurate structural modulation \cite{n6pp}.

The ordering of the excess oxygen, however, complicates the
analysis of the magnetic order and of the corresponding
excitations. Therefore, the larger part of neutron and x-ray
scattering studies was performed on \lsnod ~single crystals
\cite{n7,19,lee1,n11,n12,lee2,n14,n15,n15p,nxr1,nxr1p,nxr2,nxr3,nxr4,nxr5}.
In an oxygen-stoichiometric sample with x = 0.2 Tranquada et al.
find the magnetic correlation at \ves $\sim$ (0.125, 0.125, $l$)
peaking at odd $l$ \cite{n7}, but the modulation perpendicular to
the planes remains quite weak. Charge scattering appears at
\vech$\sim$(0.25, 0.25, $l$) again with odd $l$ but is better
defined along the $c$ direction. Note that the incommensurability
is slightly higher than expected for the amount of charges. For a
crystal with the same Sr content but less oxygen Hayden et al.
found an incommensurability of only $\eps _{S} \sim$ 0.08
suggesting $\eps _{ch} \sim$ 0.16 \cite{n2} slightly below the
expected value. These two measurements illustrate the uncertainty
arising from the oxygen concentration in these materials, but they
both prove that the stripe order of charges and spins in \lsnod \
sets in for samples which still exhibit the structural distortion
due to the octahedron tilting, T$_{HTT/LTO}\sim$70\ K in \lsnopz \
\cite{n7}. When scanned with energy analysis, elastic magnetic
correlations appear below T $\sim$ 90\ K in \lsnopz \ \cite{n7}
which is significantly below the charge ordering temperature,
$T_{co}$$\sim$115\ K , in this material. This indicates that
magnetic correlations become quasi-static on the time-scale of
this experiment below T $\sim$ 90\ K (i.e. relaxation times are
larger than a few ps). When scanning in an energy-integrating
neutron-scattering mode, which is possible due to the 2D character
of the signal, the magnetic signal in \lsnopz \ persists to higher
temperature and seems to disappear together with the charge
ordering. Charge order thus immediately triggers strong 2D
magnetic correlations in \lsno \ but these remain inelastic till
much lower temperature.  The importance of time scales is further
illustrated in a NMR study on \lsnodd \ which finds a 30K lower
magnetic transition temperature than the neutron experiments due
to the $\mu$s time scale of the NMR experiment \cite{nmr}.
Clearly, the magnetic transitions in \lsnod \ posses a glassy
character.

\begin{figure} \centering {
\includegraphics[width=0.72\columnwidth]{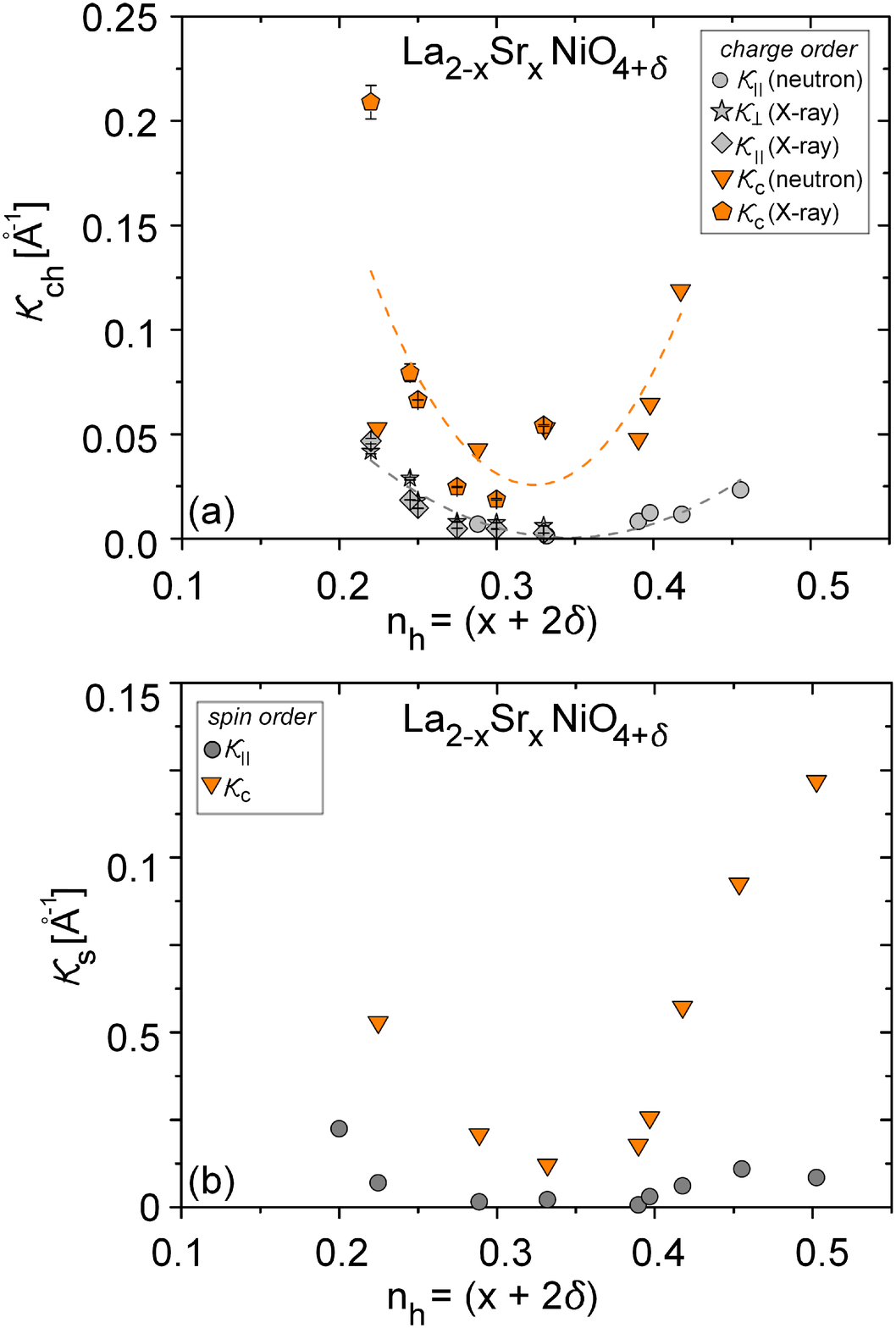}
\caption{ The inverse charge-correlation lengths along the stripe
modulation in-plane, $\kappa _\parallel ^{ch}$, perpendicular to
the modulation but parallel to the planes $\kappa _\perp ^{ch}$
(note that this direction is parallel to the stripes), and
perpendicular to the planes, $\kappa _c ^{ch}$ (above). The
inverse magnetic correlation lengths are given below: parallel to
the modulation, $\kappa _\parallel ^{S}$, and perpendicular to the
planes, $\kappa _c ^{S}$. Data are taken from references
\cite{n12,nxr3}. } \label{fig4}}
\end{figure}

The Sr concentration of x = 0.2 is the lowest one, where stripe ordering has been unambiguously established \cite{n2} yielding the smallest
modulation reported so far ($\eps _{S} \sim 0.08$). For the Sr content of x = 0.135 Tranquada et al. find static or quasistatic response at the
commensurate position of the nn antiferromagnetism \cite{n7} suggesting that \lsno \ with x = 0.135 exhibits  glassy commensurate
antiferromagnetism at a rather low T$_N$ of 65\ K. Also a study of the magnetic excitations for x = 0.12 showed the renormalized magnon
dispersion expected for a commensurate structure \cite{drack,unp-k}.

\begin{figure*} \centering {
\includegraphics[width=1.0\textwidth]{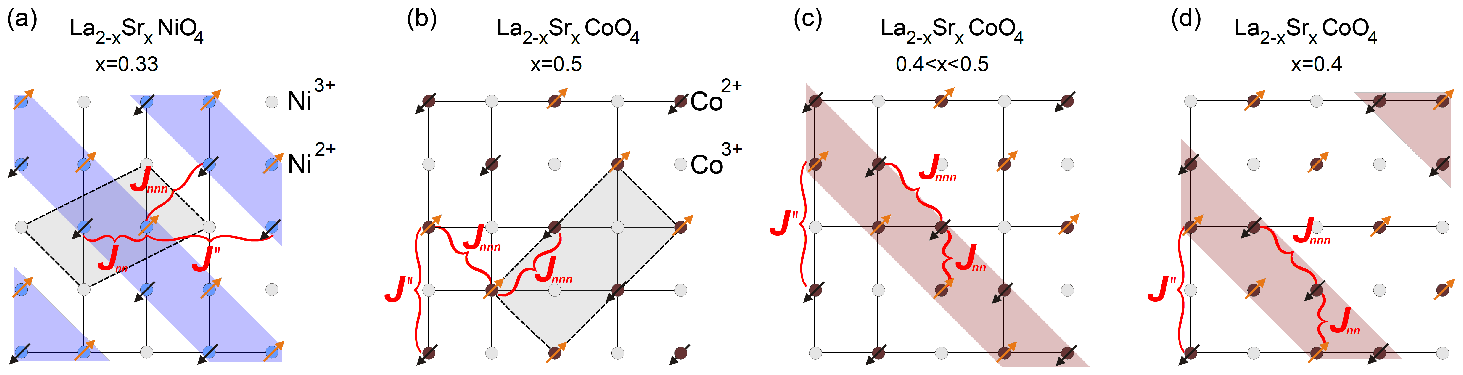}
\caption{ Scheme of the robust magnetic order occurring at x =
$1/3$ doping in the nickelates (a); magnetic interaction
parameters are indicated, $J_{nn}$ between neighboring Ni$^{2+}$;
$J_{nnn}$ and $J''$ describe the interaction across the charge
stripe. (b)-(d)  Scheme of the magnetic and charge ordering
appearing in  \lscov. (b) ideal ordering at halfdoping \lscopf ;
charges order into a checkerboard arrangement with magnetic
Co$^{2+}$ and non-magnetic Co$^{3+}$ sites; magnetic interaction
parameters between magnetic Co$^{2+}$ are indicated in the same
notation as in part (a); (c) stripe-like modulation of the
ordering slightly below half doping, where additional magnetic
stripes stabilize magnetic ordering and (d) stripe order order for
x=0.4.} \label{fig5}}
\end{figure*}

Systematic studies on the doping dependence of the stripe order in
\lsnod \ were performed by comprehensive neutron
\cite{n12,n14,n15p} and x-ray \cite{nxr2,nxr3,nxr4,n15,nxr5}
scattering studies. The resulting doping dependencies of the
transition temperatures, of the incommensurabilities and of the
inverse correlation lengths are shown in Figures
\ref{fig2},\ref{fig3} and \ref{fig4}, respectively. These studies
document an enhanced stability of the stripe phase at x = 1/3. For
this concentration the magnetic and charge ordering transition are
the highest and the correlation lengths are the longest. Just the
checkerboard charge ordering for half doping appears at even
higher temperature but this phase is again commensurate. Evidence
for the particular stability of the stripe phase at x = 1/3 is
also found in the clear specific heat anomaly \cite{ng2p}. The
magnetic structure of the x = 1/3 stripe phase is commensurate and
quite simple as the magnetic cell comprises only two antiparallel
magnetic moments, see Fig. \ref{fig5} (a).

The incommensurability for x = 1/3 fits exactly the Sr content and
does not vary with temperature. In addition, the in-plane
correlation lengths are pretty large. For example the magnetic
correlation length along the modulation vector amounts to $\xi
_\parallel ^S$ = 100 \AA \ and that of the longitudinal charge
modulation to $\xi _\parallel ^{ch}$ = 350 \AA \ while the
modulations perpendicular to the planes are still limited and
approximately equal  $\xi _c ^S\sim \xi _c ^{ch}\sim$30 \AA \
\cite{lee1}. With synchrotron radiation the in-plane anisotropy of
the charge superstructure was analyzed finding shorter correlation
lengths along the stripe modulation vector  than perpendicular to
it, $\xi _\parallel ^{ch} < \xi _\perp ^{ch}$ for Sr
concentrations between x = 0.225 and x = 0.33
\cite{nxr1,nxr1p,nxr2}, see Fig. \ref{fig4}. This anisotropy seems
to be the direct consequence of the solitonic arrangement of the
stripes. Locally, the distance between two Ni-centered charge
stripes can only amount to an integer multiple but this number has
to vary in order to adapt for the non-commensurate charge
concentration generating some diffuse scattering. In contrast,
along the stripes and thus perpendicular to the stripe modulation
the order can be perfect in principle \cite{c14}.


We note that charge-order superstructure scattering in \lsno \
always peaks at ($\eps$, $\eps$, $l$) with odd $l$, whereas the
magnetic scattering exhibits peaks for odd and even $l$
\cite{lee1,n14}. Due to the body centering of the small tetragonal
cell of the K$_2$NiF$_4$ structure in I4/mmm, (0, 0, 1) is not a
Bragg point and the propagation vectors ($\eps$, $\eps$, $l$) with
odd and even $l$ are not equivalent but refer to 180$^\circ$ or
zero phase shift for the modulation of the two Ni sites at (0, 0,
0) and at (0.5, 0.5, 0.5), respectively. The Coulomb forces
require always anti-phase stacking of the charge ordering in
agreement with the finite dispersion of phonons perpendicular to
the planes \cite{nphon1,nphon2}. In contrast, both in-phase and
anti-phase magnetic stacking is possible. For x = 0.333 only a
small amount of the even-$l$ scattering is observed \cite{lee1},
but the relative weight of the even-$l$ signals increases upon
increase or reduction of the Sr content away from x = 1/3
\cite{n14} underlining the particular stability of the charge and
magnetic order at this composition.

The relevance of the commensurate and stable x = $1/3$ structure
becomes further illustrated when looking at the temperature
dependence of incommensurabilites \cite{n15,n15p,nxr2}. The charge
incommensurability is not constant but varies upon cooling as
already observed earlier, and the comprehensive analyzes show that
the temperature dependence is determined through the doping. For
x$<$0.33, $\eps _{ch}$ is much closer to 1/3 at high temperatures
than the doping suggests but $\eps _{ch}$ decreases on cooling
approaching the expected value at low temperature; these
low-temperature values are given in Fig. \ref{fig2}. For larger
doping, x$>$1/3, the same behavior is observed but it results in
an increase of $\eps _{ch}$ upon cooling \cite{n15,n15p}. This
behavior points to an electronic self doping of the charge stripes
at higher temperature \cite{n15,n15p}. One may thus consider the
variation of charge content around x$\sim$$n_h$=$1/3$ as doping
into the stable antiferromagnetic x = $\frac{1}{3}$ Mott state
shown in Fig. \ref{fig2}. This picture is further corroborated by
the Hall effect measurements across this amount of doping which
show a change in the sign of charge carriers just at x = 1/3 and
suggest electron- and hole-like charge carriers below and above
this value, respectively \cite{ng7}. The stability of the
commensurate phase at x = 1/3 thus explains the slight deviations
of the $\eps _{ch}$$ = $$n_h$ rule that persist even at low
temperatures, see Fig. \ref{fig2}, but the overall correspondence
between charge concentration and incommensurability yields a
strong argument in favor of the stripe picture with localized
charges and moments.

There is a discrepancy between the synchrotron and the neutron
studies concerning the low-temperature behavior of the charge
ordering. Upon cooling synchrotron experiments find  a reduction
of the peak height \cite{nxr2,nxr3,nxr5}, which is accompanied by
a broadening and a slight shift of the signal. The variation in
the integrated intensity, therefore, is less dramatic. This
behavior may arise from the freezing of the correlations into the
intrinsic disorder of the materials combined with the different
time- and spatial scales of neutron and synchrotron diffraction
experiments but requires further attention.

Astonishingly little is known about the microscopic structure of
the charge order in \lsno . The average crystal structure in
\lsnodd \ seems not to change when entering the charge-ordered
phase \cite{n8,wu} but the distortions arising from the charge
modulation have not been reported so far. This limits the
reliability of density-functional theory (DFT) analyzes which need
to determine the microscopic distortions by structure relaxation
\cite{nt1,nt2,nt3}. For an oxygen doped material, NMR experiments
document the solitonic character and indicate that charge ordering
is centered at the Ni sites with S = $\frac{1}{2}$ Ni$^{3+}$
$3d^7$ ions in the domain walls \cite{n13,n13p}. Resonant
diffraction experiments using soft x-ray for \lsnopz \ reveal that
the holes are mainly located on oxygens surrounding a nominal
Ni$^{3+}$ and that the difference in electron counting is quite
small, $\Delta Z\sim0.3$ electron charges \cite{nxr4}. The
magnetic structure in the stripe phase was studied by polarized
and unpolarized-neutron diffraction revealing a spin reorientation
transition within the $ab$ plane for x = 0.275, 0.37, and 0.4
\cite{n16}.

Various inelastic neutron scattering experiments were performed to
study the magnon dispersion in pure \lno \ and in several stripe
phases \cite{nsw1,nsw2,nsw3,nsw4,nsw5,nsw6,drack}. Pure \lno \
exhibits a steep magnon dispersion with a spin-wave velocity of
340 meV\AA \ \cite{nsw1}. In a stripe phase with an oxygen excess
of $\delta$ = 0.133, spin-wave-like excitations still exhibit a
steep dispersion emanating at the incommensurate magnetic zone
centers with a spin-wave velocity reduced by 40 \% compared to
that in pure \lno \ \cite{nsw2}. Batista et al. proposed first
that the magnon dispersion of a stripe phase can explain the
peculiar resonance mode observed in the cuprates superconductors
\cite{nsw7}. In a stripe phase, magnon branches emanate at four
incommensurate spots surrounding the (0.5, 0.5, 0) position of the
nn commensurate antiferromagnetism; the merging of these
contributions should result in a resonance-like intensity
enhancement \cite{nsw7}. The magnon dispersion of the stripe
phases in \lsno \ was therefore studied in great detail focusing
on the concentration range near x = $1/3$
\cite{nsw3,nsw4,nsw5,nsw6}. For x = 0.31 close to the most stable
order a steep dispersion was observed with an almost isotropic
spin-wave velocity of about 320 meV\AA \ which is comparable to
that in pure \lno \ \cite{nsw3}. This indicates that the magnetic
coupling across the stripe  ($J_{nnn}$ and $J''$ in Fig.
\ref{fig5}) is quite strong in this simple magnetic structure.
Further experiments show that essentially the same high-energy
dispersion is observed when the doping is slightly varied
\cite{nsw3,nsw4,nsw5,nsw6}. These findings agree with the already
mentioned dominance of the x = $1/3$ phase in \lsno which seems to
prohibit the emergence of a cuprate-like dispersion of magnetic
excitations. Just half-doped La$_{1.5}$Sr$_{0.5}$NiO$_4$ with
checkerboard arrangement exhibits a different dispersion of
magnetic excitations \cite{n20,nhd2}. The S = $\frac{1}{2}$
Ni$^{3+}$ ions in the domain walls can be considered as forming a
chain-like magnetic arrangement and the characteristic features of
such a chain indeed were found in the inelastic neutron scattering
response \cite{nsw5,nsw8}.

Stable stripe order was also found in {\it Re}$_{1.67}${\it A}$_{0.33}$NiO$_4$ with {\it Re} a rare-earth element and {\it A} = Ca or Sr
\cite{n17,n18,n19}. The smaller ionic radius of the rare earth stabilizes the structural tilt distortion which pins the direction of the stripes
along the direction of the octahedral tilt axis. The disorder induced by the local variation of the ionic radii at the $Re$ site in {\it
Re}$_{1.67}${\it A}$_{0.33}$NiO$_4$  reduces the correlation lengths yielding the longest correlations for Pr$_{1.67}$Ca$_{0.33}$NiO$_4$
\cite{n19}. These findings on materials with smaller ions occupying the $Re$ site corroborate the conclusion in \lsno \ that stripe order can
coexist with the tilt distortions.

Phonons in \lsnod \ have been studied \cite{nphon1,nphon2} as
reference to the cuprates high-temperature superconductors which
exhibit strong phonon anomalies
\cite{phon4,phon4p,phon4pp,phon5,phon6,phon6a,phon7,phon8,phon8a,phon8b,phon26,reznik}.
The lattice dynamics in the undoped parent compounds can be very
well understood by ionic  lattice-dynamics model \cite{chaplot},
but in electronically doped cuprates, significant discrepancies
appear in the dispersion of the longitudinal bond-stretching
branches
\cite{phon4,phon4p,phon4pp,phon5,phon6,phon6a,phon7,phon8,phon8a,phon8b,phon26,reznik}.
Including electronic screening to the models describing the
lattice dynamics of the cuprates parent compounds yields an
increasing dispersion for the longitudinal bond-stretching
branches, since the screening is less perfect at small distances
corresponding to larger phonon propagation vectors. Doped cuprates
indeed show this behavior for several polar branches, but the
longitudinal bond-stretching branches behave differently. They
exhibit a nearly flat dispersion along [110] and even a decreasing
dispersion along [100]. In  the [100] direction a strongly
renormalized half-breathing mode is observed together with strong
frequency broadening. This anomaly can be identified with the
tendency towards stripe ordering \cite{phon6,reznik,bradenphon1}.
The phonon dispersion in doped nickelates is also anomalous,
because longitudinal bond-stretching branches soften in the
Brillouin zone \cite{nphon1,nphon2} so that the longitudinal
breathing modes fall below the transverse bond-stretching modes,
which can be considered as an overscreening effect. However, in
contrast to the cuprates, the down-bending is not limited to the
[100] direction but appears also for [110] where it is even
stronger. This corroborates the general interpretation that the
bond-stretching modes in transition-metal oxides dynamically
reflect the charge ordering instabilities \cite{bradenphon1},
because the charge-stripe modulation in nickelates appears along
the [110] direction as well. The question whether there is a more
direct signature of the charge stripe ordering in the phonon
dispersion in \lsnod \ requires further attention \cite{drack}.

\begin{figure} \centering {
\includegraphics[width=0.85\columnwidth]{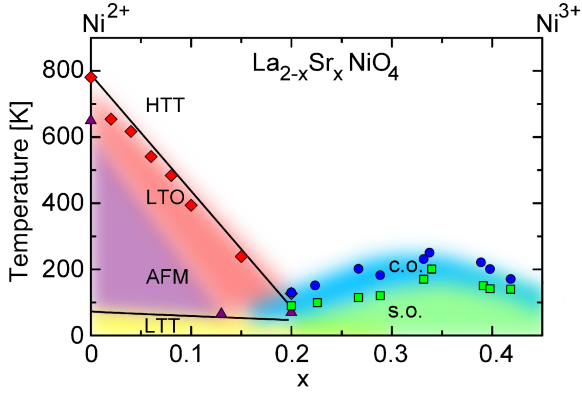}
\caption{Phase diagram of \lsno \ concerning the structural
distortion (red triangles and line), the suppression of
commensurate antiferromagnetism (magenta triangles) and the
appearance of the stripe phase with charge (blue) and magnetic
(green) transitions. The LTT or LTLO phase is indicated in yellow.
Data are taken from \cite{n7,n12,friedt}. } \label{fig6}}
\end{figure}

The schematic phase diagram of \lsno \ is shown in Fig.
\ref{fig6}; the doping suppresses both the structural distortion
and the commensurate nn antiferromagnetic order giving rise to the
charge and spin stripe ordering. The crossover from the
commensurate nn antiferromagnetism to the stripe order occurs near
$x\sim 0.2$ close to the full suppression of the structural
distortion, but there is overlap of the octahedron tilt distortion
and the stripe phases \cite{n7}.

\section{Incommensurate magnetic order in layered cobaltates}
\label{cobaltates}


In view of the enormous amount of literature on \lsno \ it is astonishing that the isostructural cobaltates, \lscoo , were only little studied.
Similar to \lsno , layered cobaltates do not become metallic unless a high amount of doping above x$\sim$1 is introduced \cite{c1,c1p,c2,c3}. In
the cobaltates one has to distinguish different possibilities to distribute the electrons in the $t_{2g}$ and $e_g$ crystal-field levels in an
octahedral arrangement. Co$^{2+}$ with seven electrons in the $d$ shell always exhibits a high-spin (HS) S = $\frac{3}{2}$ state in an
octahedral arrangement due to Hund's rule coupling \cite{c4}. In contrast, Co$^{3+}$ with six electrons in the $d$ shell can adopt different
spin states depending on the number of electrons transferred to the higher $e_g$ level: a low-spin (LS) non-magnetic state with all electrons in
the $t_{2g}$ level, an intermediate-spin (IS) state with S = 1 for one electron transferred to the $e_g$ orbital, and a HS state with S = 2 for
two transferred electrons. It is well known that the spin state of Co$^{3+}$ may even vary as a function of temperature \cite{lacoo3}. The
possible spin-state transitions as a function of doping render the analysis of the phase diagram of \lscoo \ more difficult and different
spin-state scenarios were proposed \cite{c1,hollmann}. Similar to the nickelates, the layered cobaltates also tend to incorporate large amounts
of excess oxygen rendering the growth of single crystals particularly difficult in the low-doping range \cite{c5,c6,c7,c8} whereas powder
samples can be more easily controlled \cite{c1,haider}.

Pure \lcoo \ exhibits the HTT to LTO transition at a high temperature of 900\ K \cite{kajitani} due to further increased bond-length mismatch.
Upon cooling the second structural transition to the LTT phase is found at 135\ K, and nn antiferromagnetic order develops below 275\ K
\cite{c4}. The larger structural distortion in the cobaltates is reflected in a larger amount of Sr-doping needed to fully suppress the tilting
of the octahedra: about 40\% in the cobaltates versus slightly above 20\% in the nickelates and cuprates. The single-layer magnetic structure of
pure \lcoo, \lno , and \lco \ are identical but the direction of the ordered moment in the orthorhombic structure and the stacking of the
magnetic layers is different. Adopting space group Bmab for the LTO phase, the spins point along the $a$ direction for \lcoo \ and \lno ; note
that the $a$ direction is the tilt axis in Bmab. In \lco \ spins align along $b$ perpendicular to the tilt axis, so that Dzyaloshinski-Moriya
interaction gives rise to a small canting of magnetic moments along $c$ \cite{thio}. The stacking of magnetic order in all three compounds is
described by the (1, 0, 0) vector which yields different magnetic structures due to the rotation of moments between \lco \ and \lno / \lcoo .
The two types of structure are characteristic for many compounds with K$_2$NiF$_4$ structure and may even coexist \cite{braden2}.

At half-doping, \lscopf \ exhibits checkerboard charge ordering at
the exceptionally high transition temperature of T$_{CO}$ =
825(27)\ K. In this very stable charge ordered phase magnetic
order develops only below T$_{S}$$\sim$30\ K \cite{c9,c10}. The
large ratio of T$_{CO}$/T$_{S}$$\sim$30 indicates an effective
decoupling of the two phenomena and thus implies that in this
cobaltate the charge order arises from electron-lattice coupling
\cite{c10}.

\begin{figure} \centering {
\includegraphics[width=0.75\columnwidth]{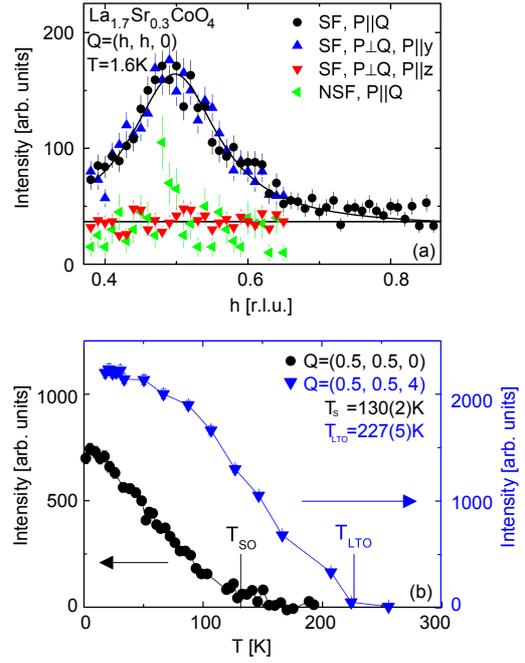}
\caption{Magnetic ordering occurring in \lscopd \ \cite{20}; scans
with neutron-polarization analysis are shown in part (a)
documenting the magnetic nature of the very broad commensurate
signal at low temperatures; the temperature dependencies of the
structural and magnetic superstructure intensities are given in
part (b).} \label{fig7}}
\end{figure}

The spin state of Co$^{3+}$ in the \lscoo \ series has generated a lot of controversy following the initial analysis of Moritomo et al.
proposing a HS to IS transition near x = 0.7 \cite{c1}. The magnetic structure analysis in \lscopf \ indicates an effectively non-magnetic
Co$^{3+}$ and a rough study of the structural distortions was interpreted as evidence for an IS state. The IS state with a single electron in
the $e_g$ shell is strongly Jahn-Teller active, and an octahedron elongation at the Co$^{3+}$ sites was interpreted as evidence for the IS state
\cite{c10}. A study of the structural distortion with a four-circle diffractometer collecting many more reflections \cite{drcwik} does not
support the IS state. Using this structural model, DFT finds a low-spin state \cite{hua}. Also the anisotropy of the magnetic susceptibility
\cite{hollmann,hollmann2} points to a low-spin state of Co$^{3+}$ in the doping range 0.4$<$x$<$0.8. In view of the large amount of conflicting
interpretations a final conclusion about the spin-state near half doping seems not yet possible
\cite{cw30,cw29,cw31,horigane1,horigane2,rogerprl}, in particular one may wonder whether simple local models can capture all the physics of
\lscoo .

\begin{figure} \centering {
\includegraphics[width=0.70\columnwidth]{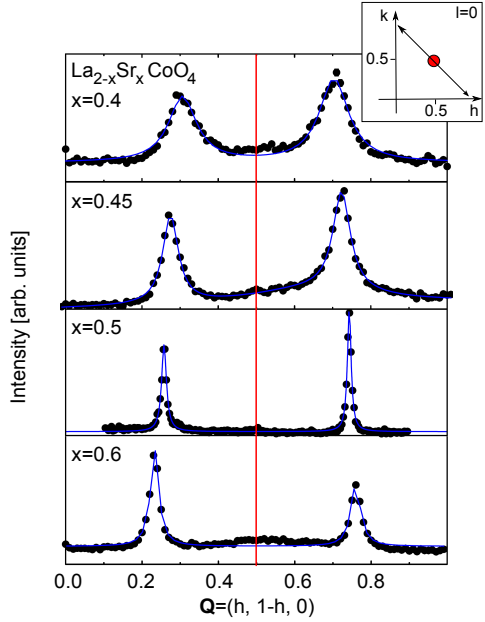}
\caption{Elastic neutron scattering scans aiming at the magnetic
ordering occurring in \lscoo ; the scans connect two satellites
depicted in Fig. 1(c) with the position of nn commensurate
antiferromagnetism. Data taken from reference \cite{20}. }
\label{fig8}}
\end{figure}

For the magnetic structure at half doping, there is no indication
for a contribution of a magnetic Co$^{3+}$ neither in the elastic
scattering \cite{c9,c10,20,drcwik} nor in the analysis of the
magnetic excitations \cite{c11,c12,drcwik}. The magnetic structure
in \lscopf \ is not perfectly commensurate but the magnetic signal
is slightly displaced already suggesting some charge
inhomogeneity. Assuming non-magnetic Co$^{3+}$ the magnetic
structure arises from the exchange between two Co$^{2+}$ in a
distance of about 8 \AA, see Fig. \ref{fig5} (b), whereas the
interaction across the diagonals is frustrated. This splits the
Co$^{2+}$-sites into two sets which can couple only through weaker
inter-layer parameters. The magnetic structure may therefore
become instable against any minor modification lifting this
degeneracy \cite{20}. An introduction of an additional row of
Co$^{2+}$ or just polarizing magnetic sites in formal Co$^{3+}$
rows lifts the degeneracy and significantly strengthens the
magnetic order, see Fig. \ref{fig5} (c).

\begin{figure} \centering {
\includegraphics[width=0.70\columnwidth]{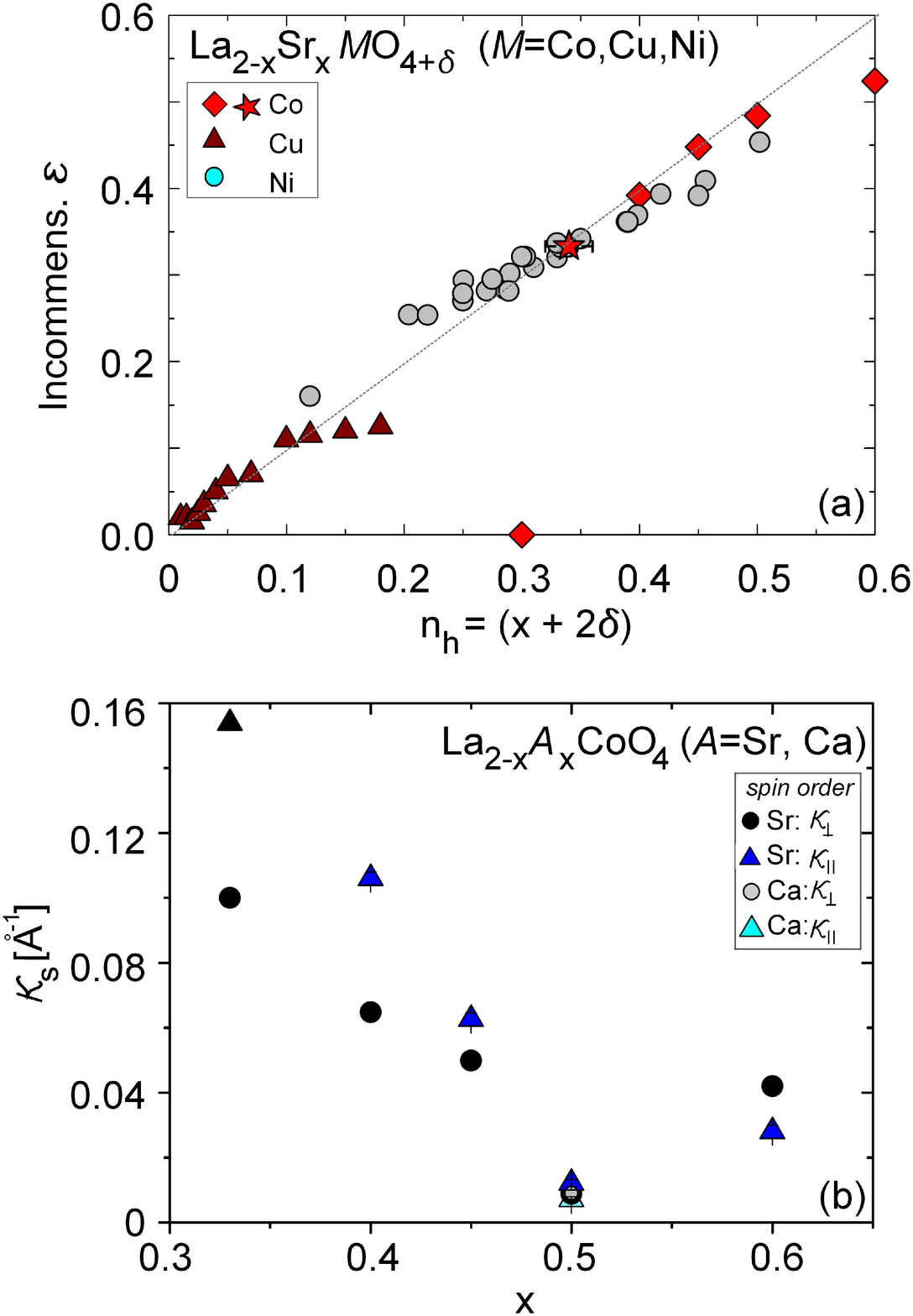}
\caption{(a) Dependence of the charge incommensurability on the
amount of doping in \lscoo \ compared to nickelates and
cobaltates; note that the values for the cuprates were scaled by a
factor two to consider the different occupation of charges in the
stripe (data taken from references \cite{20,c13}). Inverse
correlation lengths of the magnetic ordering in \lscoo \ plotted
against the amount of doping (b). Data taken from references
\cite{20,c13}. } \label{fig9}}
\end{figure}

A single crystal with x = 0.3 doping was studied by polarized- and
by unpolarized-neutron diffraction \cite{20} revealing a
significant suppression of the structural transition to $T_S$ =
227\ K, see Fig. \ref{fig7}(a). The polarization analysis further
reveals commensurate quasi-2D magnetic scattering around {\bf Q} =
(0.5, 0.5, 0) documenting that this material is still governed by
the simple nn antiferromagnetic instability. Scanning in different
directions with good resolution excludes that the broad signal
actually arises from an incommensurate stripe phase corresponding
to a doping of x = 0.3 \cite{drcwik}. The stability of the
commensurate antiferromagnetism in \lscoo \ is remarkable in view
of the rapid suppression in cuprates and in nickelates. A possible
explanation can be given in terms of the reduced mobility of
charge carriers in layered cobaltates \cite{c1} which can be
related not only with enhanced electron-phonon coupling but also
with the spin states. Maignan et al. examined the possible
electron transfer between a Co$^{2+}$ and a neighboring Co$^{3+}$
ion for different spin states in HoBaCo$_2$O$_{5.5}$, proposing
the so-called spin-blockade mechanism \cite{cw102}. For the case
of a HS Co$^{2+}$ neighboring a LS Co$^{3+}$, the $e_g$ electrons
of Co$^{2+}$ cannot hop into the free $e_g$ orbitals of Co$^{3+}$,
since this would result in two unfavorable IS states and thus cost
a high amount of energy. Electrons in a Co$^{2+}$ HS and Co$^{3+}$
LS lattice are thus effectively trapped explaining the higher
resistivity of layered cobaltates. The observed reduction of the
nn antiferromagnetic $T_N$ in \lscoo \ can be compared to the
reduction of T$_N$ in K$_2$(Co$_{1-x}$Mg$_x$)F$_4$ where a
non-mobile non-magnetic impurity is introduced
\cite{cw276,drcwik,20}. The relative suppression of T$_N$ is
identical in both systems.

\begin{figure} \centering {
\includegraphics[width=0.9\columnwidth]{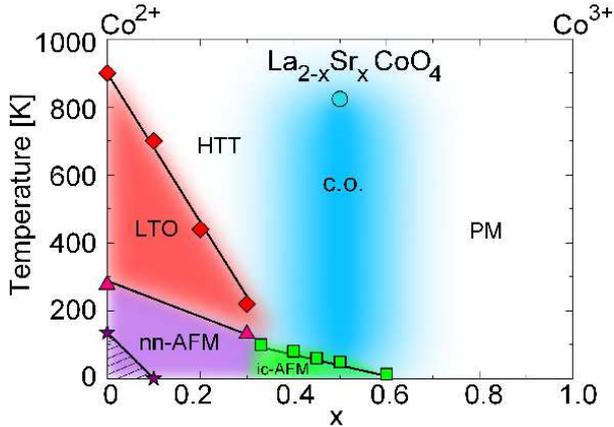}
\caption{Phase diagram of \lscoo \ indicating the structural phase transition into the LTO phase (red diamonds and line), the nn commensurate
antiferromagnetism (magenta triangles) and the incommensurate magnetic stripe order (green squares); data are taken from references
\cite{20,c13,c9,c10}. } \label{fig10}}
\end{figure}

For slightly higher Sr content incommensurate scattering emerges
similar to the case of the layered nickelates strongly suggesting
a stripe arrangement of the diagonal type [Fig. \ref{fig1}(b) and
Fig. 5(d)] \cite{20,c13}. Figure \ref{fig8} shows scans across the
commensurate center of nn antiferromagnetism for x = 0.4, 0.45,
0.5, and 0.6 (data taken from reference \cite{20}). Figure
\ref{fig9}(b) presents the doping dependence of the low
temperature inverse correlation lengths and of the
incommensurabilities found in these experiments. The magnetic
scattering in \lscoo \ remains quite broad and exhibits a
pronounced in-plane anisotropy. The correlation along the
modulation is much shorter than that perpendicular to it, which
qualitatively agrees with the findings in the nickelates and the
expectation for a solitonic stripe arrangement \cite{c14}. For the
higher doping, the distance between two charges stripes is very
short in this picture. Also the minimum of the inverse correlation
length at x = 0.5, see Fig. \ref{fig9}, resembles the behavior in
\lsno , see Fig. \ref{fig4}, but in the nickelates, maximum
stability appears already at x = $1/3$. The most stable ordering
in the cobaltates also manifests itself by larger correlation
lengths and a higher integrated intensity. The magnetic phase is
found to be more stable in \lccopf \ due to the structural tilt
distortion in this material and due to the reduced $A$-site
disorder similar to the analysis of {\it Re}$_{1.67}${\it
A}$_{0.33}$NiO$_4$ \cite{n19}.

The doping dependence of the incommensurability agrees almost
perfectly with the expectation for the stripe ordering  for
$0.3\le x$$\le$0.5 and it naturally fits into the relations
obtained for cuprates and nickelates suggesting the common
mechanism in terms of charge segregation, see Fig. \ref{fig9}(a).
Note that $\eps_{ch}$ has been scaled by a factor two in the case
of the cuprates in order to take the different occupation in the
charge stripe into account. Compared to cuprates and nickelates,
stripe order in \lscoo \ just starts at higher doping
concentration.  The incommensurability $\eps_{ch}$ for
La$_{1.4}$Sr$_{0.6}$CoO$_4$ falls well below the expected value
indicating the limit of the fully localized stripe picture in the
cobaltates.

There is, however, only weak evidence for incommensurate charge
ordering in \lscoo \ just a very diffuse signal was observed.
Incommensurate magnetism has been reported also for higher doping
but the underlying structure is not fully clarified \cite{c15}.

The resulting phase diagram for \lscoo \ closely resembles that
for \lsno , see Fig. \ref{fig10} and Fig. \ref{fig6}. The
structural distortion and the commensurate antiferromagnetism are
suppressed with the doping. The slower suppression of the
octahedron tilting can be attributed to the higher bond-length
mismatch of the pure compound. The more robust nn
antiferromagnetic structure has been proposed to result from the
spin-blockade mechanism \cite{20}. The crossover from commensurate
to incommensurate antiferromagnetism appears again close to the
full suppression of the structural distortion but still in the
distorted phase. The second observation of an almost coincidence
of the suppression of the structural distortion with the onset of
stripe order eventually indicates some implication of the tilt
disorder in the emergence of the stripe phase. The mixing of La
and Sr on the same site induces significant disorder in \lscoo \
that is further enhanced by the softening of the tilt instability
around the structural transition. The influence of the La-Sr
mixing on local tilt distortions was analyzed for \lsco
\cite{braden-disorder} and should be even stronger in the case of
the cobaltates due to the larger amount of Sr doping. Due to the
disorder, one may not expect a sharp structural transition neither
as function of doping nor as function of temperature, but strong
local distortions should exist on both sides of the transition.
These might favor the stripe order and further destabilize the
commensurate antiferromagnetism. The magnetic transition
temperatures presented in Fig. \ref{fig10} have to be regarded
with caution, because they most likely are subject to severe
time-scale issues. The data in reference \cite{20} were measured
on a cold instrument yielding a characteristic time scale of a few
$p$s, any fluctuation slower than this will appear as an elastic
signal in this experiment. The decrease of magnetic transition
temperatures appears to be almost linear in the doping, even
across the transition from commensurate to incommensurate
magnetism, which may reflect and thus support the continuous
suppression of magnetic moments with doping as expected for LS
Co$^{3+}$.

Magnetic excitations in \lscoo \ were studied for pure \lcoo \
\cite{c16}, for half-doped \lscopf \ \cite{c11,c12,drcwik} and
more recently also for \lscoz \ \cite{c13}. The magnon dispersion
in the pure and in the half-doped material can be well described
by linear spin-wave theory and thus allows one to determine the
magnetic interaction between two neighboring Co$^{2+}$ moments
$J_{nn}$ and that between two Co$^{2+}$ moments connected by an
intermediate non-magnetic Co$^{3+}$, $J''$ see Fig. \ref{fig5} (d)
\cite{c13}. In \lscoz \ Boothroyd et al. find magnetic stripe
ordering very similar to the stable state in \lsnodd \ \cite{c13}
but with very short in-plane correlation lengths of only
$\xi_{\parallel}^S$ = 6.5 \AA \ and $\xi_{\perp}^S$ = 10 \AA \
parallel and perpendicular to the stripe modulation, respectively
(note that these directions are perpendicular and parallel to the
stripes, respectively). Most interestingly they find the magnon
dispersion to exhibit an hourglass-shaped dispersion, like it has
been observed for many cuprates materials \cite{13,14,15,16,17}.
This strengthens the interpretation, that fluctuating stripes are
the proper basis to describe the magnetic excitation in the
cuprates in contrast to purely itinerant models \cite{18}. Phonon
studies have not been reported for layered cobaltates, but one may
expect similar anomalies as in the case of the nickelates
\cite{nphon1,nphon2}.

\begin{figure} \centering {
\includegraphics[width=0.75\columnwidth]{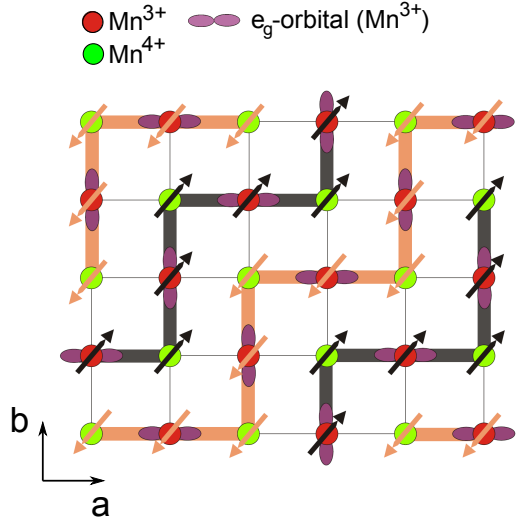}
\caption{Schematic arrangement of the charge, orbital, and
magnetic ordering in half doped manganites, a single layer is
shown. In the Goodenough model charges exhibit checkerboard order
with orbital order occurring on the \mnd \ $3d^4$ sites with a
single $e_g$ electron. The aligned $e_g$ orbital mediates a strong
ferromagnetic interaction resulting in zigzag ferromagnetic chains
that are antiferromagnetically stacked. } \label{fig11}}
\end{figure}

\section{Incommensurate order of charges, spins and orbitals in layered manganites}
\label{cobaltates}

The colossal magnetoresistance (CMR) reported for several
manganese oxides \cite{tokura00a,tokura} stimulated comprehensive
studies in experiment and theory, since this phenomenon promised a
high potential for applications. The Zener double-exchange
mechanism can only explain a small part of the drop of resistivity
by several orders of magnitude that is implied by only moderate
external magnetic fields \cite{millis95a}. The essential part of
the CMR effect arises from the competition between insulating
antiferromagnetic states and a ferromagnetic metallic phase. By
application of a magnetic field one induces an insulator-metal
transition into the ferromagnetic state which is accompanied by
the loss of resistivity \cite{tokura00a,murakami03a}.
Understanding the CMR thus means understanding the interplay of
the different phases including their segregation
\cite{uehara99a,moreo99a,woodward04a,milward05a,sen07a} and, most
importantly, understanding the character of the insulating phases.
These insulating phases exhibit charge and orbital ordering
accompanied by antiferromagnetic ordering at low temperatures, but
the phase diagrams of the manganites exhibit a remarkable variety
of such charge and orbital ordered phases \cite{tokura}. In spite
of comprehensive experimental and theoretical efforts the
knowledge about the insulating state is still limited and
controversial proposals have been made even for fundamental
questions like  site \cite{m1,m1p,m1pp} or bond centering
\cite{m2} of the charge order. The role of the stripe order for
high-temperature superconductivity can still be questioned, but
there is no doubt that charge-ordered phases are essential for the
CMR effect.

\begin{figure} \centering {
\includegraphics[width=0.8\columnwidth]{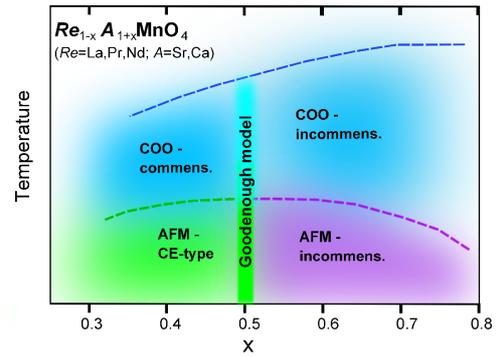}
\caption{Phase diagram of \lsmov \ indicating the charge, orbital
and magnetic ordering near half doping; note that phase diagrams
of various perovskite phases are very similar. Underdoping with
respect to half doping results in an almost commensurate order
very similar to the Goodenough model described in Fig.
\ref{fig11}, whereas overdoping immediately implies incommensurate
modulations that perfectly follow the amount of doping.}
\label{fig12}}
\end{figure}

The most stable and best characterized charge and orbital ordered
phase occurs at half doping either in the perovskite materials,
($Re_{1-x}A_x$MnO$_3$ with $Re$ = La, Y or a rare earth and $A$ in
most cases an earth alkali) or in single- or double-layered
compounds. The model proposed very early for
La$_{0.5}$Ca$_{0.5}$MnO$_3$ consists of a checkerboard ordering of
the charges accompanied by orbital ordering at the \mnd \ site and
is called Goodenough model, it is depicted in Fig. \ref{fig11}
\cite{m1p,m1pp}. \mnd \ ion with $3d^4$ configuration and a single
electron in the $e_g$ shell is strongly Jahn-Teller active and the
aligning of the orbital mediates a strong ferromagnetic
interaction with the neighboring \mnv \ moments (corresponding to
a $3d^3$ configuration). In Fig. \ref{fig11}, we show a schematic
picture of this charge, orbital, and magnetic arrangement for a
single MnO$_2$ layer. Due to strong structural distortions even
the perovskite materials exhibit some layered character with the
planes of charge and orbital order always aligning along the $ab$
planes (in space group Pbnm). Single-layered manganites \lsmov \
\cite{m3} do not exhibit the CMR effect unless very high magnetic
fields are applied and even then only bad metallic properties are
observed \cite{m4}. However, these materials are much better
suited for any scattering studies of charge and orbital order due
to the absence of the complex twining and due to the availability
of large single crystals \cite{reutler}. Half-doped \lsmohd \ has
been studied by many different techniques
\cite{m5,m6,m7,m8,m9,m10,m11,m12,m13} and the charge and orbital
ordered state of this material can be considered as the best
characterized one amongst all manganites. It is well established
for this material, that the Goodenough model is the appropriate
one to describe the charge and orbital order \cite{m12,m13} but
only qualitatively. The modulation of the electronic charge is far
below the integer value but the orbital polarization at the
nominal \mnd \ site is nearly complete \cite{m11}. The small
modulation of the electronic charge is also found in perovskite
materials \cite{m14,m15,m16} so that it can be considered as
representative for charge ordered states in manganites. Four
different types of superstructure reflections can be identified in
half-doped \lsmohd \ \cite{m16,m18,21}. Neglecting the inter-layer
coupling, charge order causes weak super structure reflections at
{\bf k}$_{\text{ch}}$
 = $\pm$(0.5, 0.5). The orbital order can be considered as the main element of the structural distortion in \lsmohd  \ \cite{m7,m11} and is
associated with the strongest nuclear superstructure peaks
appearing at {\bf k}$_{\text{oo}}$ = $\pm$(0.25, 0.25). The
orbital arrangement can be considered as stacking of rows of
parallel orbitals with a 90\grad \ rotation of the orbital order
in neighboring orbital rows. The magnetic order is described by
two propagation vectors {\bf k}$_{\text{Mn}^{3+}}$ = $\pm$(0.25,
-0.25) and {\bf k}$_{\text{Mn}^{4+}}$=$\pm$(0.5, 0.0) referring to
the nominal \mnd \ and \mnv \ spins, respectively.

The phase diagram of \lsmov \ is quite different to those
discussed in the previous sections, as we start for x = 0 with
LaSrMnO$_4$ which only possesses \mnd \ sites and is thus
comparable to LaMnO$_3$. The transport, magnetic \cite{m3} and
structural \cite{m17} properties of this system are well known.
The pure material exhibits nn antiferromagnetic ordering with the
moments aligning along the $c$ direction. This magnetic structure
is coupled with an $e_g$ orbital alignment along the $c$ direction
\cite{m10,m17}. Upon further increase of the Sr content the
antiferromagnetism is suppressed and orbitals flop into the planes
as is indicated in Fig. \ref{fig11} for x = 0.5 \cite{m17}.

\begin{figure} \centering {
\includegraphics[width=0.8\columnwidth]{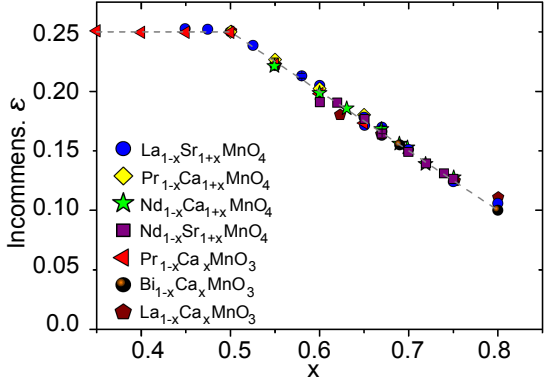}
\caption{Dependence of the orbital modulation on the doping across
the half-doped composition for various manganite systems. Below
half doping the order stays almost commensurate, but above half
doping the orbital incommensurability perfectly follows the amount
of excess \mnv \ with respect to half doping, $\eps_{oo}$ =
0.25-${\frac{\Delta x}{2}}$. This strict rule underlines the
dominant role of the orbital ordering in the stripe phases in
overdoped manganites. Data were taken from references
\cite{m10,21,norimatsu-la} for La$_{1-x}$Sr$_{1+x}$MnO$_4$,
\cite{mm15,hour} for Pr$_{1-x}$Ca$_{1+x}$MnO$_4$, \cite{mm15b} for
Nd$_{1-x}$Ca$_{1+x}$MnO$_4$, \cite{mm9,norimatsu-la-b} for
Nd$_{1-x}$Sr$_{1+x}$MnO$_4$, \cite{mm15} for
Pr$_{1-x}$Ca$_{x}$MnO$_3$, \cite{grenier} for
Bi$_{1-x}$Ca$_{x}$MnO$_3$, and \cite{23} for
La$_{1-x}$Ca$_{x}$MnO$_3$.} \label{fig13}}
\end{figure}

Evidence for an incommensurate ordering of charges and orbitals
has been found in electron diffraction experiments shortly after
the revival of interest in La$_{1-x}$Ca$_x$MnO$_3$ due to the
discovery of the CMR effect \cite{22,23,23a}. The first
experiments were interpreted in terms of simple charge ordering
and revealed a linear relationship between the incommensurability
and the electronic doping, which should be considered as the
finger print of a stripe phase \cite{22,23,23a}. In our notation,
the strongest nuclear superstructure reflections appear at \veo =
($\eps_{oo}$, $\eps_{oo}$, $q_l$) shifted from a fundamental Bragg
peak and arise from orbital ordering \cite{m6,m13}. The relation
$\eps_{oo}$ = 0.5-${\frac{x}{2}}$ = 0.25-${\frac{\Delta x}{2}}$
with $\Delta x$ the amount of overdoping,  $\Delta x$ = $x$-0.5,
is perfectly fulfilled in many different manganites systems
including perovskite and layered phases
\cite{22,23,m6,m10,21,hour,mm15,norimatsu-la,grenier,norimatsu-la-b,mm15},
see Fig. \ref{fig13}. There is a striking asymmetry between over-
and underdoping with respect to half doping: Slightly below x =
0.5 the orbital order appears at or close to the commensurate
quarter-indexed position, whereas the linear relation is
immediately fulfilled  for even small overdoping.

The perfect confirmation of the $\eps_{oo} = 0.25-{\frac{\Delta
x}{2}}$ rule for so many different manganites systems with
perovskite and layered structure is remarkable
\cite{22,23,m6,m10,21,hour,mm15,norimatsu-la,grenier,norimatsu-la-b,mm15},
see Fig. \ref{fig13}. For overdoping with respect to halfdoping,
the linear relation is almost perfectly fulfilled for
La$_{1-x}$Sr$_{1+x}$MnO$_4$  \cite{m10,21,norimatsu-la},
Pr$_{1-x}$Ca$_{1+x}$MnO$_4$ \cite{mm15,hour},
Nd$_{1-x}$Ca$_{1+x}$MnO$_4$ \cite{mm15b},
Nd$_{1-x}$Sr$_{1+x}$MnO$_4$ \cite{mm9,norimatsu-la-b},
Pr$_{1-x}$Ca$_{x}$MnO$_3$ \cite{mm15} , Bi$_{1-x}$Ca$_{x}$MnO$_3$
\cite{grenier}, Bi$_{1-x}$Sr$_{x}$MnO$_3$ \cite{mm26}, and
 La$_{1-x}$Ca$_{x}$MnO$_3$ \cite{23}. The underlying orbital order certainly is the key element to
understand the stripe phases in overdoped manganites. The law can
be easily explained with the Wigner model: There is strong
tendency to form diagonal rows of parallel \mnd \ orbitals with
the distance between them given by the amount of excess \mnv . A
further factor two in the incommensurability arises from the
90$^\circ$ rotation of the orbitals in neighboring orbital
stripes. As for the CE-type ordering at half doping, one may then
ask whether this orbital ordering has a purely structural or a
partially magnetic mechanism
\cite{solovyev99a,brink99a,solovyev03a,solovyev01a,brey05a,daghofer06a}.
The ordering in a single orbital stripe implies a strong
ferromagnetic interaction of the central \mnd \ with the two \mnv
\ neighboring moments along the occupied $e_g$ orbital, in
addition the stacking of these ferromagnetic trimers is
antiferromagnetic.

We wish to emphasize, however, that the stripe phase in overdoped
manganites is fundamentally different from all those discussed
above, because it is already a charge and orbitally ordered state
which becomes incommensurately modulated by the variation of
doping. The asymmetry is most likely related with the fact that
\mnd \ can take the place of \mnv \ by flopping the orbital
perpendicular to the plane \cite{mm28}.

In spite of strong efforts
\cite{22,23,25,mm8,mm9,mm10,mm11,mm11p,milward05a,mm13,mm14,mm15,mm16,mm17,mm15b,mm25,m10,m11,mm24,mm26,pissa,pissa1,grenier}
the incommensurate stripe-like phases for overdoping are
insufficiently understood and essential questions about the nature
of the electronic arrangement and about the homogeneity of the
modulation remain matter of controversy. The observation of the
superstructure reflections in the diagonal direction indicates
orbital rows running along the diagonals. The excess rows of \mnv
\ with respect to half doping run along the diagonals as well.
However, there are different possibilities to arrange the orbital
rows. In the so-called Wigner model the rows with parallel \mnd \
orbitals arrange as far as possible, whereas high-resolution
transmission electron microscopy suggests that two such \mnd \
rows together with a shared \mnv \ line form a bi-stripe
\cite{25}. The bi-stripe can be identified with a single "zig" in
the zigzag ferromagnetic chains of the CE-type structure, see Fig.
\ref{fig11},  at half-doping; this model suggests a strong
stability of the bi-stripe local distortion. Also a double stripe
of two directly neighboring \mnd \ rows was proposed \cite{mm26}.

Numerous electron, x-ray and neutron diffraction experiments
studied the stripe phases in various overdoped manganites,
however, without reaching a clear conclusion \
\cite{mm8,mm9,mm10,mm13,mm14,m10,m11,pissa,pissa1}. Concerning the
question about bi-stripe versus Wigner model, the larger part of
the experiments favor the Wigner model
\cite{mm25,mm24,mm8,grenier} for the perovskites compounds,
whereas experiments on double-layer manganites were interpreted in
terms of the bi-stripe model \cite{mm13,mm14} but a clear
experimental proof for one of the two scenarios is still missing.
Note, however, that only for doping above x = 0.6 there is a
difference between the two models as otherwise bi-stripes form
also in the Wigner model.

Both a soliton lattice \cite{mm11,mm11p,milward05a,21,mm17} and a
homogenous charge-density wave \cite{mm10,mm16} have been proposed
to explain the incommensurabilities generating again strong
controversy. However, the recent observation of remarkably strong
second-order harmonics of the orbital ordering reflection as well
as NMR experiments give strong support for a solitonic arrangment
of well defined stripes of excess \mnv \ \cite{21,mm17}, whereas
the argumentation in reference \cite{mm10} can be questioned.

The interplay between the orbital order and the magnetism is well
studied for \lsmohd \ \cite{m12,m13}. Both phenomena are closely
coupled, because the $e_g$ orbital always implies a strong
directed ferromagnetic interaction. Therefore, it is not easy to
analyze which instability is driving and which is following
\cite{solovyev99a,brink99a,solovyev03a,solovyev01a,brey05a,daghofer06a}.
The sequence of the magnetic correlations with isotropic
short-range ferromagnetic correlations above T$_{co}$ and
anisotropic ferromagnetic and antiferromagnetic correlations for
T$_N$$<$T$<$T$_{co}$ and long-range CE-type ordering below T$_N$
is obtained in several analyzes
\cite{solovyev03a,brey05a,daghofer06a}. The fact that the orbital
scattering coexists at high temperature T$>$T$_{co}$ with
ferromagnetic short-range order suggests strong electron-lattice
coupling driving the orbital ordering. This conclusion is further
corroborated by measurements on overdoped \lsmovo \ \cite{21},
\pcmovz , and \nsmovz \ \cite{hour} in which incommensurate
orbital scattering coexists with ferromagnetic correlations
without any measurable change in the position.

\begin{figure*} \centering
\includegraphics[width=1.8\columnwidth]{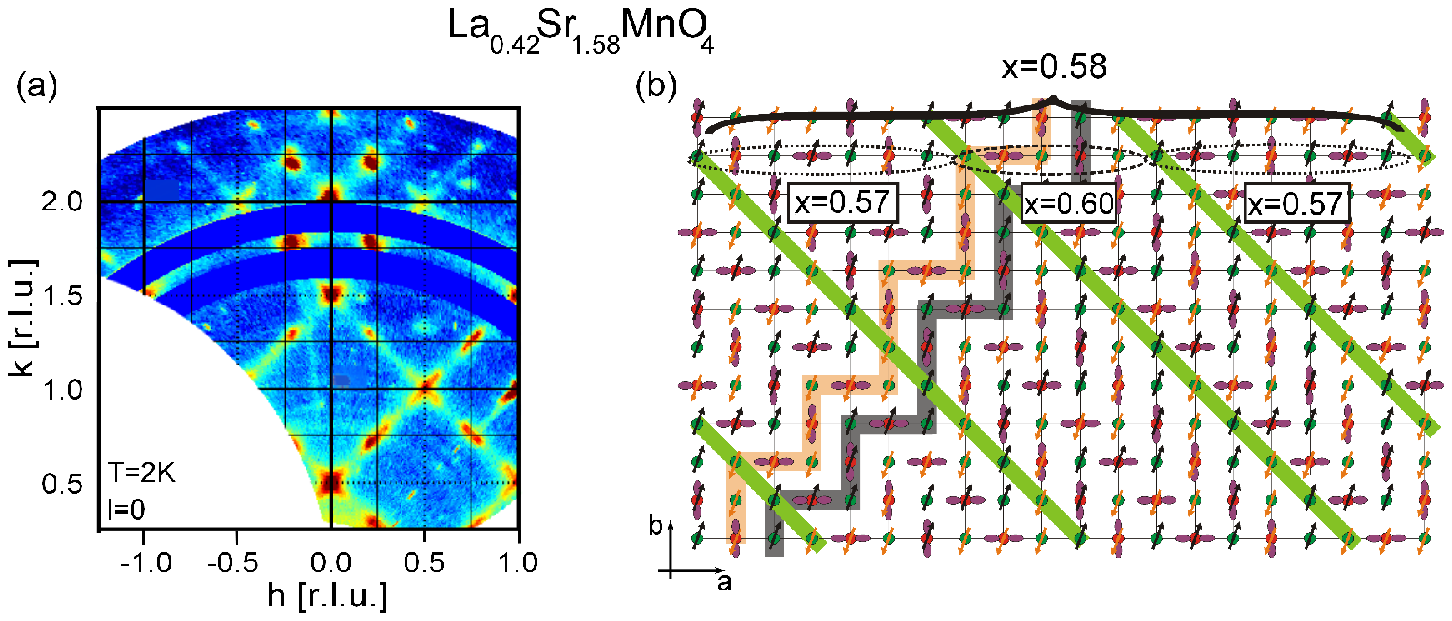}
\caption{Distribution of structural and magnetic scattering in the
(h,k,0) plane for  \lsmovo \ arising from incommensurate ordering
of charge, orbital and magnetic ordering of {\mnd} (a). Sketch of
charge, orbital and spin order for \lsmovo \ (b). Red circles
represent {\mnd} and green circles {\mnv}. A single domain of
zigzag chains propagating in [110] direction is shown. The excess
lines of {\mnv} are displayed on a green background. A local
variation of stripe distances is needed to account for the real
charge concentration (data taken from references
\cite{m10,23,mm15,mm15b,21,hour}).} \label{fig14}
\end{figure*}

Due to the complex twining in the perovskites and due to the
complex arrangement of the orbital, charge and spin order in
overdoped manganites, a full model of the different order
parameters could not be developed for the perovskite phases. For
slightly  overdoped layered manganites, \lsmovo \ and
La$_{0.45}$Sr$_{1.55}$MnO$_4$, a full mapping of the neutron
scattering was measured as function of temperature and allowed
establishing a consistent model \cite{21}. Fig. \ref{fig14} shows
the low temperature map together with a model for the stripe
ordering in this material. The mapping as well as comprehensive
additional neutron diffraction experiments indicate that the sharp
superstructure peaks in \lsmohd \ become incommensurately
displaced and broadened in \lsmovo . Three incommensurate and one
commensurate superstructure can be identified. Charge order
reflections are found at {\bf Q} =
(1.5$\pm$\text{$\eps_{\text{ch}}$}
1.5$\pm$\text{$\eps_{\text{ch}}$} 0) with
\text{$\eps_{\text{ch}}$} = 0.080(3). The orbital satellites are
centered closer to the Bragg-reflection, for example {\bf Q} = (0,
2, 0), in diagonal direction in perfect agreement with data for
layered and perovskite manganites, {\bf k}$_{\text{oo}}$ =
(0.25-$\Delta\eps_{\text{oo}}$,0.25-$\Delta\eps_{\text{oo}}$,0).
Neither the charge nor the orbital scattering exhibit any
measurable temperature dependency concerning their positions, and
the incommensurabilities of charge and orbital scattering differ
precisely by a factor 2, ($\Delta\eps_{oo}$ = 0.039(2) =
$\frac{1}{2}\eps_{ch}$).

The magnetic scattering associated with the \mnd \ is also
incommensurate in \lsmov \ with x slightly above 0.5. The
modulation is transverse with respect to the quarter-indexed peaks
in \lsmohd , {\bf k}$_{\text{Mn}^{3+}}$ =
(0.25-$\eps_{\text{Mn}^{3+}}$, 0.25+$\eps_{\text{Mn}^{3+}}$, 0)
\cite{21}, and the modulation length exactly agrees with the two
nuclear ones: $\eps_{\text{Mn}^{3+}}$ = 0.037(2) = $\Delta
\eps_{oo}$ = $\frac{1}{2}\eps_{\text{ch}}$. In contrast to these
three incommensurate order parameters, the magnetic scattering
associated with the ordering of {\mnv} spins remains commensurate.

The close coupling of the structural and magnetic superstructures
corroborates the stripe-like arrangement previously deduced from
the structural distortions only. Since all modulations point along
the diagonals, the excess \mnv \ must align along the diagonals
thereby interrupting the ferromagnetic zigzag chains of the
CE-type structure which run along the perpendicular diagonals. A
model for the charge orbital and stripe order in \lsmovo \ is
presented in Fig. \ref{fig14}. For x = 0.6 a simple commensurate
model consists of a bi-stripe combined with a double \mnv \
stripe. This pattern has a width of five diagonal rows. For x =
0.57$\sim$$\frac{4}{7}$ a similar commensurate model is
constructed with a width of seven rows. In order to obtain an
incommensurate arrangement corresponding to the doping of x =
0.58, the five- and seven-rows models can be combined in the
appropriate ratio as it is illustrated Fig. \ref{fig14}. This
solitonic arrangement of stripes gives a perfect description of
all scattering intensities in \lsmovo \ \cite{21}. In particular
the simple Fourier transform of large supercells describes the
diffuse shape of the magnetic scattering which exhibits
characteristic tails along the diagonals. This shape can be taken
as further evidence for the solitonic stripe arrangement
\cite{c14}.

\begin{figure} \centering {
\includegraphics[width=0.75\columnwidth]{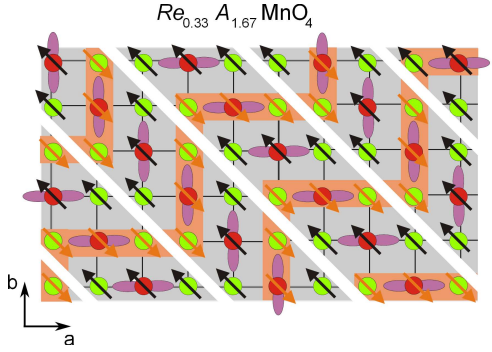}
\caption{Schematic picture of the charge orbital and magnetic
ordering in x = 2/3 doped layered manganites,
Nd$_{0.33}$Sr$_{1.67}$MnO$_4$ and Pr$_{0.33}$Ca$_{1.67}$MnO$_4$;
the main structural element consists in rows of \mnd \ sites with
parallel $e_g$ orbitals, that lead to a magnetically well defined
block of three Mn sites width. The magnetic coupling between these
orbital stripes, however, is frustrated.} \label{fig15}}
\end{figure}

Increasing the doping beyond x = 0.6 one may distinguish between
the Wigner and the bi-stripe arrangement of the orbital ordering.
The perfect relation between the incommensurability and the
doping, which shows no anomaly near x = 0.6, indicates that the
orbital ordering continues smoothly across this value. Clearly,
orbital ordering is the main element to describe the complex
ordered states in overdoped manganites. Although we still lack a
full experimental proof, there is stronger support for the Wigner
model in perovskite and single-layered manganites in the doping
rang 0.6$<$x$<$0.8 \cite{mm25,mm24,mm8,grenier}. Mapping the
different superstructure intensities in
Nd$_{0.33}$Sr$_{1.67}$MnO$_4$ and in Pr$_{0.33}$Ca$_{1.67}$MnO$_4$
with the flatcone neutron detector, we also find clear evidence
for the Wigner model in the single-layered manganites \cite{hour}.
Orbital ordering in heavily overdoped manganites consists of
diagonal stripes of \mnd \ with parallel alignment of the
$e_g$-orbitals. These orbital rows result in three Mn-sites thick
blocks with strong magnetic coupling, see Fig. \ref{fig15}. The
two \mnv \ moments neighboring the \mnd \ are ferromagnetically
coupled and there is an antiferromagnetic coupling in between two
such trimers. However, the coupling between two stripes is very
weak due to frustration, see Fig. \ref{fig15}. The magnetic
scattering is quite different to that appearing at lower
overdoping. Instead of incommensurate \mnd \ and commensurate \mnv
\ magnetic scattering, one finds incommensurate signals around the
\mnv \ position and commensurate quarter-indexed scattering for x
= 2/3. The fact that the incommensurate magnetic modulations
change between and x = 0.58 and 0.67 without any anomaly in the
orbital ordering, see Fig. \ref{fig13}, clearly indicates that the
orbital pattern is independent of the magnetic arrangement. The
orbital pattern is mainly associated with the magnetic coupling
within the stripe whereas the inter-stripe coupling represents a
much smaller energy scale, see below.

There is also evidence for self-organized charge, orbital and
magnetic ordering in some underdoped manganites \cite{ye}, but
different propagation vectors seem to coexist and a conclusive
model of all types of order for such materials is still lacking.

There are numerous studies on the magnetic excitations focussing
on the metallic ferromagnetic phases, see the discussion in
references \cite{fm1,fm2}, and on the low-doping \cite{petit}
concentration range, but the magnon dispersion in the CE-type
phase at half doping has been reported only recently
\cite{m12,m13,m18} due to the difficulties to analyze this complex
order with twinned crystals. In the CE-type phase, the strongest
magnetic coupling is found between a \mnd \ and \mnv \ moment
coupled through the ordered $e_g$ orbital, like it is expected in
the Goodenough model. This dominating magnetic interaction gives
rise to strong anisotropy in the spin-wave dispersion, which is
much steeper along the ferromagnetic zigzag chains
\cite{m12,m13,m18}. Very recently, the spin-wave dispersion has
also been studied for overdoped manganites which exhibit a stripe
arrangement of orbitals and charges. Most interestingly,
Nd$_{0.33}$Sr$_{1.67}$MnO$_4$ and Pr$_{0.33}$Ca$_{1.67}$MnO$_4$
exhibit a hourglass dispersion very similar to those reported for
the cuprates \cite{13,14,15,16,17} and for \lscoz \ \cite{c13}.
Pr$_{0.33}$Ca$_{1.67}$MnO$_4$ exhibits a better defined magnetic
ordering at low temperature due to structural distortions that the
smaller A-site ions imply, and due to the smaller internal
disorder. Note that Pr and Ca possess similar ionic radii, whereas
those of La and Sr or of Nd and Sr differ considerably. The longer
correlation lengths in Pr$_{0.33}$Ca$_{1.67}$MnO$_4$ offer an
additional insight in the causes for the hourglass-shaped
dispersion. At low temperature, when the correlation lengths are
longer, one finds the outwards dispersing branches but these
become suppressed upon heating when correlation lengths diminish.
Nd$_{0.33}$Sr$_{1.67}$MnO$_4$ exhibits short correlation lengths
even at low temperature and no indication for the outwards bending
branches \cite{hour}. Short magnetic correlation lengths appear to
be a necessary condition for the insulating stripe phases to
exhibit an hourglass dispersion. The short correlation lengths in
x = 2/3 doped manganites are a consequence of the frustrated
coupling between the magnetically well determined orbital stripes,
see Fig. \ref{fig15}. In this view these manganites closely
resemble the cuprates, with weak coupling of the antiferromagnetic
blocks across the charge stripes, and also the layered cobaltate,
where the non-magnetic Co$^{3+}$ sites reduce the coupling across
the charge stripes.

\section{Conclusions}

Overwhelming evidence for stripe-like arrangement of charges,
magnetic moments and orbital degrees of freedom has been found in
several layered non-cuprates transition-metal oxides.
Incommensurate ordering of charges, orbitals, and magnetic moments
should be considered as the rule and not as the exception in
transition-metal oxides. The main difference with the stripes
phases discussed for the cuprates concerns the insulating
properties of the non-cuprates compounds, which are supposed to
arise from a stronger electron-lattice interaction. There is
reasonable evidence that the stripe patterns with solitonic and
local moment character depicted in Fig. \ref{fig1} and \ref{fig14}
are better suited to describe these insulating phases than a
coupled CDW/SDW scenario, whereas this question still is an open
issue in the cuprates.

In none of the systems described here, one finds an integer
modulation of the charge at the metal ion site
\cite{coey,m11,m14,m15,m16}. Like in other charge ordered
transition-metal oxides, e.g. Fe$_3$O$_4$, a purely ionic model is
insufficient to describe the charge and orbital ordering even for
these insulating materials. Due to covalency and hybridization
with the oxygen orbitals a larger part of the charge modulation in
all these materials occurs on the bonds surrounding the metal ion,
but the ordering seems to be still centered at the metal site in
the layered nickelates and manganites. This spread charge
modulation, however, does not contradict the local character of
the stripe phase depicted.

The phase diagrams of La$_{2-x}$Sr$_x$$M$O$_4$ with $M$ = Cu, Ni
or Co are remarkably similar. All the pure materials exhibit a
structural phase transition characterized by octahedron tilting
that is followed by a second structural transition due to a change
of the tilt axis in the case of Ni and Co. This structural
instability is perfectly explained through the bond-length
mismatch, which is partially released by the doping due to the
larger ionic radius of Sr and due to the smaller induced radius at
the oxidized metal ion. Due to the stronger starting mismatch in
\lcoo \ it takes a larger amount of doping to fully suppress the
structural distortion in \lscoo \ compared to \lsno \ and \lsco .
Also the nn antiferromagnetism is suppressed with the doping in
all three systems, but the amount of doping needed for full
suppression does not correlate with the $T_N$ of the parent
compound. The mobility of the induced holes seems to be decisive
for the suppression of the nn antiferromagnetism, occurring at
$\sim$2\% , $\sim$17.5\% \ and $\sim$30\% \, in cuprates,
nickelates and cobaltates, respectively. All three systems show
evidence for magnetic stripe ordering appearing next to the
commensurate phases. In nickelates and cobaltates the onset of
stripe order almost coincides with the suppression of the tilt
distortion. Due to the intrinsic disorder induced by the doping
the structural transition is not well-defined in \lsno \ and in
\lscoo \ and the enhanced local variation of the tilt angles in
the vicinity of the transition can help stabilizing the stripe
order or destabilizing the nn commensurate antiferromagnetic
order.

The relationship between the structural and magnetic modulation
and the amount of charge doping, $\eps_{ch}\propto n_h$, can be
considered as the characteristic finger print of a stripe phase
with local charge and magnetic character, as a CDW/SDW instability
arising from Fermi-surface features can adopt any value. Ti-doped
Sr$_2$RuO$_4$ is an example of an isostructural metallic compound
that exhibits incommensurate magnetic ordering arising from a
nesting instability \cite{tisrruo,sidis} which should be
considered as a homogenous SDW.  The $\eps_{ch}\propto n_h$
relationship is better fulfilled in the insulating materials than
in the cuprates, although there still are deviations. In the
nickelates the very stable order at $1/3$ doping deforms the
$\eps_{ch}(x)$ relation in particular at higher temperatures.
Similar effects are found in the layered cobaltates for $x>0.5$
and in the manganites for $x < 0.5$, associated with an asymmetry
of the phase diagram around the stable phase at half doping.
Self-doping and non-local effects thus are already present in
these insulating stripe phases: at both sides of the 1/3 doping in
the case of nickelates, whereas cobaltates and manganites tend to
assimilate the stable ordering only at one unfavorable side.

In the manganites, clear evidence for stripe ordering with
localized character is found in materials that are overdoped with
respect to half-doping. These phases and the resulting phase
diagrams are different, because it is an already charge and
orbital ordered state that becomes further modulated or
interrupted by rows of additional charges. The orbital ordering
clearly constitutes the most important element to understand the
complex ordering in the manganites. For large overdoping \mnd \
sites and their $e_g$ orbitals form stripes, and the distance
between them is perfectly well described by the amount of doping
(assuming a Wigner model). The perfectly fulfilled
$\eps$=0.25-$\frac{\Delta x}{2}$ relation in many different
families of overdoped manganites demonstrates the outstanding
stability of the orbital ordering with the strong tendency to form
an orbital stripe which then mediates the magnetic coupling. There
are indications in the manganites that the orbital and the
incommensurate magnetic ordering are decoupled.

The analysis of the magnetic excitations in the stripe-ordered
insulating phases is most interesting with respect to the
hourglass-shaped dispersion reported for the cuprates. Stripe
ordered nickelates do not exhibit such a dispersion due to the
very stable order at $1/3$ doping with a strong interaction across
the charge stripes. However \lscoz ,
Nd$_{0.33}$Sr$_{1.67}$MnO$_4$, and Pr$_{0.33}$Ca$_{1.67}$MnO$_4$
were recently shown to exhibit the characteristic features of the
hourglass dispersion. The hourglass dispersion differs from that
predicted by linear spin-wave theory mostly by the suppression of
the outwards dispersing branches. This feature was shown to be
caused by the reduction of the magnetic correlation lengths to the
order of the distance between the charge stripes. The common
element of the metallic cuprates and the insulating cobaltates and
manganites seems to consist in the local magnetic structure
arising from well defined magnetic stripes that remain loosely
coupled. In the cuprates and in the cobaltates these stripes
consist of the nn antiferromagnetic regions in between two charge
stripes, see Fig. \ref{fig1}(b) and (c), whereas they are formed
through the orbital ordering in the manganites, see Fig.
\ref{fig15}. In all three cases the magnetic interaction between
these stripes is extremely weak reminding a magnetically smectic
phase \cite{7}. Insulating stripe patterns with local character
thus yield the same hourglass dispersion as the cuprates, however
this does not exclude a more itinerant model needed to fully
understand the cuprates.

\medskip

This work was supported by the  Deutsche Forschungsgemeinschaft
through Sonderforschungsbereich 608. We acknowledge long
collaboration and numerous stimulating discussions with M. Cwik,
A.C. Komarek, T. Lorenz, C. Sch\"ussler-Langeheine, O.J. Schumann,
D. Senff, Y. Sidis, P. Steffens, and J. Tranquada.


\end{document}